\documentclass[journal,twocolumn,10pt,twoside]{IEEEtran}
\usepackage{cuted}
\usepackage{amsfonts}
\usepackage{amsmath,cases,amssymb}
\usepackage[dvips]{graphicx}
\usepackage{cite,color}
\usepackage{color}
\usepackage{subcaption}
\usepackage{epsfig}
\usepackage{epstopdf}
\usepackage{url}
\usepackage{algorithm}
\usepackage{algorithmicx, algpseudocode}
\usepackage{color}
\usepackage{multirow}
\usepackage{mathtools}

\newcommand{\ds}{\displaystyle}

\newcommand{\clD}{{\cal D}}

\newcommand{\clM}{{\cal M}}
\newcommand{\clC}{{\cal C}}
\newcommand{\clB}{{\cal B}}
\newcommand{\clK}{{\cal K}}

\newcommand{\clR}{{\cal R}}

\newcommand{\qed}{\vspace*{-10.0mm}
\begin{flushright}
$\Box$
\end{flushright}
}

\newtheorem{myth}{Theorem}
\newtheorem{mypro}{Proposition}
\newtheorem{mylem}{Lemma}

\newcommand{\bgeqn}{\begin{equation}}
\newcommand{\edeqn}{\end{equation}}
\newcommand{\la}{\langle}
\newcommand{\ra}{\rangle}

\newcommand{\beqa}{\begin{eqnarray}}
\newcommand{\eeqa}{\end{eqnarray}}
\newcommand{\beqas}{\begin{eqnarray*}}
\newcommand{\eeqas}{\end{eqnarray*}}

\newcommand{\bK}{\mathbf{K}}
\newcommand{\bX}{\mathbf{X}}
\newcommand{\bY}{\mathbf{Y}}

\newcommand{\bW}{\mathbf{W}}

\newcommand{\bR}{\mathbf{R}}
\newcommand{\clQ}{{\cal Q}}
\begin{document}
\title{Parameterized Bilinear Matrix Inequality Techniques in ${\cal H}_{\infty}$ Fuzzy PID Control Design}
\author{Y. Shi and H. D. Tuan
\thanks{Ye Shi and Hoang Duong Tuan are with the Faculty of Engineering and Information Technology,
University of Technology, Sydney, NSW 2007, Australia. Email:
\textsf{ye.shi@student.uts.edu.au}, \textsf{tuan.hoang@uts.edu.au}.}
}
\date{}
\maketitle
\begin{abstract}
Proportional-integral-derivative (PID) structured controller is the most popular class of industrial control but still
could not be appropriately exploited in fuzzy systems.  To gain the practicability and tractability
of fuzzy systems, this paper develops a parameterized bilinear matrix inequality  characterization for the
${\cal H}_{\infty}$ fuzzy PID control design, which is then relaxed into a bilinear matrix inequality  optimization problem of
nonconvex optimization. Several computational procedures are then developed for its solution.
The merit of the developed algorithms is shown through the benchmark examples.
\end{abstract}

\begin{IEEEkeywords}
Tagaki-Sugeno (T-S) fuzzy system, $H_{\infty}$ fuzzy proportional-integral-derivative (PID) control,
parameterized bilinear matrix inequality (PBMI), bilinear matrix inequality (BMI), nonconvex optimization
techniques.
\end{IEEEkeywords}

\section{Introduction}
Tagaki-Sugeno (T-S) fuzzy model \cite{TS85} has proved as one of the most practical tools for representing complex
nonlinear systems by gain-scheduling systems, which are easily implemented online. Treating T-S fuzzy models as
gain-scheduling systems allows the application of advanced gain-scheduling control techniques in tackling
state feedback and output feedback stabilization of nonlinear systems \cite{TANY01,TANK04}. Until now, most of the
gain-scheduling controllers are assumed structure-free and full-rank to admit computationally tractable
parameterized linear matrix inequality (PLMI) or linear matrix inequality (LMI) formulations \cite{AT00,TANY01,TANK04}.

Meanwhile, proportional-integral-derivative (PID) structured  controller is the indispensable component
of industrial control so that PID control theory is still the subject of recent research
\cite{ACL05,AT03,GA11,GH15,BHA15,HTN17}, mainly concerning with linear time-invariant systems in the frequency domain.
PID controller for  fuzzy systems has been considered in \cite{GLCP15}.
Reference \cite{ZWLH01} proposed an LMI based iterative algorithm for a proportional-integral (PI) controller
in T-S systems under the  specific structure of both  system and controller.
A recent work \cite{CGLV16} transformed the fuzzy diagonal PID controller into a static output feedback problem with
the dimension of controller dramatically increased.
That is why  all its testing examples  are restricted on single input and single output systems with
two states.

This paper is concerned with the  PID parallel distribution compensation (PDC) for T-S fuzzy models. The control design
problem is formulated as a parameterized bilinear matrix inequality (PBMI) optimization problem that is in contrast to the PLMI formulation for the structure-free  PDC design \cite{TANY01}. This  is quite expected because the PID controller design for linear time-invariant systems is already nonconvex, which is equivalent to a BMI optimization problem in the state space.
In our approach, PBMI is then relaxed to a bilinear matrix inequality (BMI) for more tractable computation. It should be noted that BMI optimization constitutes one of the most computational challenging problems, for which there is no efficient computational methodology. The state-of-the-art BMI solvers \cite{ANP08,Detal12} in addressing the structure-constrained stabilizing controllers for linear time-invariant systems
must initialize from a feasible controller and then move within a convex feasibility subset containing this initialized point.  Usually their convergence is very slow \cite{Detal12}. Furthermore,
finding a feasible structure-constrained stabilizing controller is still a NP-hard problem \cite{BT97}. The most
efficient method to find such a feasible controller  is via the so-called spectral abscissa
optimization \cite{AN06}, which seeks a controller such that the state matrix of the closed loop system has
only eigenvalues with negative real parts.  This spectral abscissa  optimization-based approach cannot be extended to
gain-scheduling systems, whose stability does not quite depend on the spectrum of the time-varying state matrix. The main
contribution of the present paper is to develop efficient computational procedures for the BMI arisen from the PBMI optimization,
which generate a sequence of unstabilizing controllers that rapidly converges to the optimal stabilizing controller.

The rest of this paper is organized as follows. Section II is devoted to formulating the $H_{\infty}$ fuzzy PID control in T-S system by a PLMI, which is then relaxed by a system of BMIs.
Several nonconvex optimization techniques for addressing this BMI system are developed in Section III. Simulation for benchmark
systems is provided in Section IV to support the solution development of the previous sections. Section V concludes the paper.

{\it Notation.} Notation used in this paper is standard. Particularly, $X\succeq 0$, $X\succ 0$, $X\preceq 0$ and $X\prec 0$
mean that a symmetric matrix $X$ is positive semi-definite, positive definite, negative semi-definite and negative definite,
respectively. $\mbox{Trace}(X)$ represents the trace of $X$, while $||X||^2=\mbox{Trace}(XX^T)$ is its square norm.
In symmetric block matrices or long matrix expressions, we use $\ast$ as an ellipsis for terms that are induced
by symmetry, e.g.,
\[
K\begin{bmatrix} S+(\ast) & \ast\\ M & Q\end{bmatrix}\ast \equiv K\begin{bmatrix} S+S^T & M^T\\ M & Q\end{bmatrix}K^T.
\]
All matrix variables are boldfaced. Denote by $I_n$ the identity matrix of dimension $n\times n$ and by $0_{n\times m}$ the
zero matrix of dimension $n\times m$. The subscript $n\times m$ is omitted when it is either not important or is clear in context.
\section{$H_{\infty}$ fuzzy PID PDS for  T-S systems}
Suppose that $x$ is the state vector
with dimension $n_x$, $u$ is the control input with dimension $n_u$,
$y$ is the measurement output with dimension $n_y$, $w$ and $z$ are the disturbance and controlled output of the system with the same dimension $n_{\infty}$, and $L$ denotes the number of IF-THEN rules.
In T-S fuzzy modeling, each $i-$th plant rule is the form
\begin{equation}\label{if1}
\begin{array}{rl}
\mbox{IF} & z_1(t) \ \mbox{is} \ N_{i1} \ \mbox{and} \ \dots \ z_p(t) \ \mbox{is} \ N_{ip} \\[0.2cm]
\mbox{THEN}&
\begin{bmatrix}
\dot{x}\\z\\y
\end{bmatrix}=
\begin{bmatrix}
A_i & B_{1i} & B_{2i}\\
C_{1i} & D_{11i} & D_{12i}\\
C_{2} & D_{21} & 0
\end{bmatrix}
\begin{bmatrix}
x\\w\\u
\end{bmatrix}.
\end{array}
\end{equation}
Here $z_i$ are premise variables, which are assumed independent of the control $u$, and $N_{ij}$ are fuzzy sets. Denoting
 by $N_{ij}(z_i(t))$ the grade of membership of $z_i(t)$ in $N_{ij}$, the weight $w_i(t)=\prod_{j=1}^p N_{ij}(z_i(t))$ of each $i-$th IF-THEN rule is then normalized by
\begin{eqnarray}
\alpha_i(t)&=& \frac{w_i(t)}{\sum_{j=1}^Lw_j(t)}\geq0,\quad i=1,2,\dots,L \label{alpha1} \\
\Rightarrow\alpha(t)&=&(\alpha_1(t),\dots,\alpha_L(t))\in\Gamma\nonumber,
\end{eqnarray}
with
\begin{equation}\Gamma:=\{\alpha\in \mathbb{R}^L: \sum_{i=1}^L\alpha_i=1,\alpha_i\geq 0\}. \label{alpha2}
\end{equation}
In the state space, the T-S model is thus represented by the following gain-scheduling system
\begin{equation}\label{ts1}
\begin{bmatrix}
\dot{x}\\z\\y
\end{bmatrix}=
\begin{bmatrix}
A(\alpha(t)) & B_1(\alpha(t)) & B_2(\alpha(t))\\
C_{1}(\alpha(t)) & D_{11}(\alpha(t)) & D_{12}(\alpha(t))\\
C_{2} & D_{21}& 0
\end{bmatrix}
\begin{bmatrix}
x\\w\\u
\end{bmatrix}
\end{equation}
where
\begin{eqnarray}\label{ts2}
\begin{bmatrix}
A(\alpha(t)) & B_1(\alpha(t)) & B_2(\alpha(t))\\
C_{1}(\alpha(t)) & D_{11}(\alpha(t)) & D_{12}(\alpha(t))\\
C_{2} & D_{21} & 0
\end{bmatrix}&=& \nonumber \\
\sum_{i=1}^L\alpha_i(t)
\begin{bmatrix}
A_i & B_{1i} & B_{2i}\\
C_{1i} & D_{11i} & D_{12i}\\
C_{2} & D_{21} & 0
\end{bmatrix}.&&
\end{eqnarray}
In this paper, we seek the output feedback controller in the class of PID PDC with each $i$-th plant rule inferred by
\begin{equation}\label{if2}
\begin{array}{rl}
\mbox{IF} & z_1(t) \ \mbox{is} \ N_{i1} \ \mbox{and} \ \dots \ z_p(t) \ \mbox{is} \ N_{ip} \\[0.2cm]
\mbox{THEN}&
\begin{bmatrix}
\dot{x_K}\\u
\end{bmatrix}=
\left[
\begin{array}{cc|c}
0_{n_u\times n_u}&0_{n_u\times n_u}&\mathbf{R}_{I,i}\\
0_{n_u\times n_u}&-\tau I_{n_u}&\mathbf{R}_{D,i}\\
\hline
I_{n_u}&I_{n_u}&\mathbf{R}_{P,i}
\end{array}
\right]
\begin{bmatrix}
x_K\\y
\end{bmatrix}
\end{array}
\end{equation}
for a given $\tau>0$, where
\[
\mathbf{R}_{x,j}\in {\clR}^{n_u\times n_y}, x\in\{I,D,P\}.
\]
Note that the transfer function of this $i$-th plant rule is
\begin{eqnarray}
\bK_i(s)&=&\mathbf{R}_{P,i}+\frac{\mathbf{R}_{I,i}}{s}+\frac{\mathbf{R}_{D,i}}{s+\tau}\label{pid2}\\
&=&\mathbf{K}_{P,i}+\frac{\mathbf{K}_{I,i}}{s}+\frac{\mathbf{K}_{D,i} s}{1+\varepsilon s},\label{pid2e}
\end{eqnarray}
with $\mathbf{K}_{P,i}:= \mathbf{R}_{P,i}+\varepsilon\mathbf{R}_{D,i}$,  $\mathbf{K}_{I,i}:= \mathbf{R}_{I,i}$,
$\mathbf{K}_{D,i}:= -\varepsilon^2\mathbf{R}_{D,i}$,  and $\varepsilon=1/\tau$. It is clear from (\ref{pid2e}) that
$\mathbf{K}_{P,i}$, $\mathbf{K}_{I,i}$ and $\mathbf{K}_{D,i}$ respectively
are the proportional,  integral and  derivative
gain matrices,  while $\epsilon$ is a small tuning scalar which determines how close the last term in (\ref{if2}) comes to a pure derivative action \cite{ABN07}. In other words, (\ref{if2}) is the state-space representation of multi-input multi-output
PID structured controllers.

The PID PDC with rule set by (\ref{if2}) and the membership function defined by (\ref{alpha1})
 is thus represented by the following gain-scheduling PID controller
\begin{equation}\label{pid4}
\begin{bmatrix}
\dot{x_K}\\u
\end{bmatrix}=
(\sum_{i=1}^L\alpha_i(t)\clK_i(\mathbf{R}_i))
\begin{bmatrix}
x_K\\y
\end{bmatrix}
\end{equation}
where
\begin{eqnarray}
\clK_i(\mathbf{R}_i)=\left[
\begin{array}{cc|c}
0_{n_u\times n_u}&0_{n_u\times n_u}&\mathbf{R}_{I,i}\\
0_{n_u\times n_u}&-\tau I_{n_u}&\mathbf{R}_{D,i}\\
\hline
I_{n_u}&I_{n_u}&\mathbf{R}_{P,i}
\end{array}
\right], \label{pid5}\\
i=1,\dots, L,\nonumber
\end{eqnarray}
for
\[
\bR_i=\begin{bmatrix}
\bR_{I,i}\cr
\bR_{D,i}\cr
\bR_{P,i}
\end{bmatrix}\in \mathbb{R}^{(3n_u)\times n_y}, i=1,\dots, L.
\]
The  ${\cal H}_{\infty}$ control problem consists of finding the stabilizing controller (\ref{pid4}) for (\ref{ts1}) to solve
\begin{subequations}\label{Hi}
\begin{eqnarray}
\gamma\rightarrow \min:\gamma>0,\label{Hia}\\
\ds\int_0^T||z(t)||^2dt\leq \gamma^2\int_0^T||w(t)||^2dt \\\nonumber \quad\forall w, \quad \forall T>0, \quad x(0)=0.\label{Hib}
\end{eqnarray}
\end{subequations}
Using the shorthand
\[
\begin{array}{c}
A_{0i}=\left(\begin{matrix}A_i&0\cr
0&0\end{matrix}\right) \in \mathbb{R}^{(n_x+2n_u)\times (n_x+2n_u)},\\ B_{01i}=\left(\begin{matrix}B_{1i}\cr
0\end{matrix}\right)\in \mathbb{R}^{(n_x+2n_u)\times n_{\infty}},\\
C_{01i}=\left(\begin{matrix}C_{1i}&0\end{matrix}\right)\in \mathbb{R}^{n_{\infty}\times (n_x+2n_u)},\\
\clB_i=\left(\begin{matrix}0&B_{2i}\cr
I_{2n_u}&0\end{matrix}\right)\in \mathbb{R}^{(n_x+2n_u)\times 3n_u},\\
\clC=\left(\begin{matrix}0_{2n_u\times n_x}&I_{2n_u}\cr
C_2&0_{n_y\times 2n_u}\end{matrix}\right)\in \mathbb{R}^{(2n_u+n_y)\times (n_x+2n_u)}, \\ \clD_{12i}=\left(\begin{matrix}0&D_{12i}\end{matrix}\right)
\in \mathbb{R}^{n_{\infty}\times 3n_u},\\
\clD_{21}=\left(\begin{matrix}0\cr
D_{21}\end{matrix}\right)\in \mathbb{R}^{(2n_u+n_y)\times n_{\infty}},
 x_{cl}=\left(\begin{matrix}x\cr
x_K\end{matrix}\right),
\end{array}
\]
and then defining
\begin{eqnarray}\label{par1}
\begin{bmatrix}
A_0(\alpha) & \clB(\alpha)\\
C_{01}(\alpha) & \clD_{12}(\alpha)
\end{bmatrix}:&=&\sum_{i=1}^L\alpha_i
\begin{bmatrix}
A_{0i} & \clB_i\\
C_{01i} & \clD_{12i}
\end{bmatrix}, \nonumber \\ \clK(\alpha(t)):&=&\sum_{i=1}^L\alpha_i(t)\clK_i(\bR_i),\nonumber \\
x_{cl}&=&(x^T, x_K^T)^T \nonumber
\end{eqnarray}
the closed-loop system (\ref{ts1}), (\ref{pid4}) is rewritten by
\begin{eqnarray}
\left[\begin{matrix}\dot{x}_{cl}\cr
z\end{matrix}\right]=\left[\begin{matrix} A_0(\alpha(t))+\clB(\alpha(t))\clK(\alpha(t))\clC\cr
C_{01}(\alpha(t))+\clD_{12}(\alpha(t))\clK(\alpha(t))\clC \end{matrix}\right|\nonumber\\[0.2cm]
\left.\begin{matrix}B_{01}+\clB(\alpha(t))\clK(\alpha(t))\clD_{21}\cr
D_{11}(\alpha(t))+\clD_{12}(\alpha(t))\clK(\alpha(t))\clD_{21} \end{matrix}\right]
\left[\begin{matrix}x_{cl}\cr
w\end{matrix}\right].&\label{lct5}
\end{eqnarray}

Using the quadratic Lyapunov function $V(t):=x_{cl}^T(t)\mathbf{X}x_{cl}(t)$, $0\prec \mathbf{X}\in \mathbb{R}^{(n_x+2n_u)\times (n_x+2n_u)}$ to make (\ref{Hib}) fulfilled   by forcing
\[
\dot{V}(t)+\gamma^{-1}||z(t)||^2-\gamma||w(t)||^2\leq 0,
\]
one can easily see that (\ref{Hib}) is fulfilled by the following parameterized matrix inequality
\begin{subequations}\label{lct9}
\begin{eqnarray}
\left[\begin{matrix}
(A_0(\alpha)+\clB(\alpha)\clK(\alpha)\clC)\bX+(\ast)\cr
(B_{01}(\alpha)+\clB(\alpha)\clK(\alpha)\clD_{21})^T \cr
(C_{01}(\alpha)+\clD_{12}(\alpha)\clK(\alpha)\clC)\bX
\end{matrix}\right|\quad\quad\quad\quad\quad\quad\quad&&\nonumber\\
\left.\begin{matrix}\ast&\ast\cr
-\gamma I&\ast\cr
D_{11}(\alpha)+\clD_{12}(\alpha)\clK(\alpha)\clD_{21}&-\gamma I\end{matrix}\right]\prec 0,&&
\label{lct9a} \\[0.2cm]
\bX \succ 0, \quad\forall \alpha\in \Gamma, \label{lct9b}
\end{eqnarray}
\end{subequations}

Set
\begin{equation}\label{wj}
\mathbf{W}_i:=\clK_i(\mathbf{R}_i)\clC\mathbf{X}, i=1,\dots, L.
\end{equation}
Then
\[
\clK(\alpha)\clC\bX=\sum_{i=1}^L\alpha_i\mathbf{W}_i.
\]
For
\begin{eqnarray}
\clM_{ij}(\mathbf{X},\mathbf{R}_j,\mathbf{W}_j,\gamma)&:=&\nonumber\\[0.2cm]
\left[\begin{matrix}(A_{0i}\mathbf{X}+\clB_i\mathbf{W}_j)+\ast\\
(B_{01i}+\clB_i\clK_j(\mathbf{R}_j)\clD_{21})^T \\
C_{01i}\bX+\clD_{12i}\mathbf{W}_j
\end{matrix}\right|\quad\quad\quad\quad\quad\quad\quad&&\nonumber\\
\left.\begin{matrix}
 \ast&\ast\cr
-\gamma I&\ast\cr
D_{11i}+\clD_{12i}\clK_j(\mathbf{R}_j)\clD_{21}&-\gamma I
\end{matrix}\right], &&\label{PLMI2}
\end{eqnarray}
which is linear in its variables, the parameterized matrix inequality (\ref{lct9a}) is written by
\begin{equation}\label{paa}
\sum_{i=1}^L\sum_{j=1}^L\alpha_i\alpha_j\clM_{ij}(\mathbf{X},\mathbf{R}_j,\mathbf{W}_j,\gamma)\prec 0\quad \forall\ \alpha\in\Gamma.
\end{equation}
It follows from \cite[Theorem 2.2]{TANY01} that (\ref{paa}) is guaranteed by the following matrix inequalities
\begin{eqnarray}\label{c1}
  \clM_{ii}(\mathbf{X},\mathbf{R}_i,\mathbf{W}_i,\gamma) \prec 0, \quad i=1,2,\cdots,L\label{c1a}\\
  \frac{1}{L-1}\clM_{ii}(\mathbf{X},\mathbf{R}_i, \mathbf{W}_i,\gamma)+
  \frac{1}{2}(\clM_{ij}(\mathbf{X},\mathbf{R}_j, \mathbf{W}_j,\gamma) \nonumber \\
  +\clM_{ji}(\mathbf{X},\mathbf{R}_i,\mathbf{W}_i,\gamma))\prec 0,\label{c1b} \quad 1 \leq i \neq j \leq L.
  \end{eqnarray}
Thus the upper bound of (\ref{Hi}) is provided by the following  optimization problem:
\begin{eqnarray}
\min_{\gamma,\bX, \bR, \mathbf{W}}\ \gamma \quad\mbox{s.t.}\quad (\ref{lct9b}), (\ref{wj}),
(\ref{c1a}), (\ref{c1b}). \label{ub1a}
\end{eqnarray}
which is a BMI optimization in the decision variables $\bX$, $\mathbf{R}=(\mathbf{R}_1,\dots, \mathbf{R}_L)$ and
$\mathbf{W}=(\mathbf{W}_1,\dots,\mathbf{W}_L)$ due to the bilinear constraints (\ref{wj}).

We address this optimization problem through the following bisection procedure for a given computational
tolerance $0<\eta<<1$.

{\bf Bisection procedure.} Start from $\gamma_u$ such that the BMI system
\begin{equation}\label{bmib}
(\ref{lct9b}), (\ref{wj}),
(\ref{c1a}), (\ref{c1b})
\end{equation}
is feasible for $\gamma=\gamma_u$. Check the feasibility of BMI (\ref{bmib}) for $\gamma=(1-\eta)\gamma_u$.
If BMI (\ref{bmib}) is feasible, reset $\gamma_u=\gamma$. Otherwise, reset $\gamma_l=\gamma$. Stop until
$(\gamma_u-\gamma_l)/\gamma_u\leq \eta$ and accept $\gamma_u$ as the optimal $H_{\infty}$ gain.

The next section is devoted to address the BMI feasibility problem (\ref{bmib}). Its outcome is also a simple method to
find an initial $\gamma_u$ to start the above bisection procedure.
\section{Nonconvex spectral optimization techniques for solving BMIs}
The sparse structure of matrix $\clC$ in (\ref{wj}) suggests that (\ref{wj}) is a sparse nonlinear constraint in the sense
that there are not so many nonlinear terms in its right hand side. Indeed, by partitioning
\begin{equation}\label{p1}
\begin{array}{c}
0\prec \bX=\begin{bmatrix}\bX_{11}&\bX_{12}&\bX_{13}\cr
*&\bX_{22}&\bX_{23}\cr
*&*&\bX_{33}
\end{bmatrix},\\
\bX_{11}\in \mathbb{R}^{n_x\times n_x}, \bX_{1j}\in \mathbb{R}^{n_x\times n_u}, j=1, 2;\\
\bX_{22}\in \mathbb{R}^{n_u\times n_u}, \bX_{23}\in \mathbb{R}^{n_u\times n_u}, \bX_{33}\in \mathbb{R}^{n_u\times n_u}
\end{array}
\end{equation}
with $\bX_{ii}$ symmetric, it can be checked that
\begin{eqnarray}\label{long_2}
\clK_j(\mathbf{R}_j)\clC\bX &=&\nonumber\\[0.2cm]
\left[\begin{matrix}
\bR_{I,j}C_2\bX_{11}\cr
-\tau\bX_{13}^T+\bR_{D,j}C_2\bX_{11}\cr
\bX_{12}^T+\bX_{13}^T+\bR_{P,j}C_2X_{11}
\end{matrix}\right|\quad\quad\quad\quad\quad\quad\quad&&\nonumber\\
\left.\begin{matrix}\bR_{I,j}C_2\bX_{12} \cr
-\tau\bX_{23}^T+\bR_{D,j}C_2\bX_{12}\cr
\bX_{22}+\bX_{23}^T+\bR_{P,j}C_2\bX_{12}\end{matrix}
\right|\quad\quad\quad\quad\quad&&\nonumber\\
\left.\begin{matrix}
\bR_{I,j}C_2\bX_{13}\cr
-\tau\bX_{33}+\bR_{D,j}C_2\bX_{13}\cr
\bX_{23}+\bX_{33}+\bR_{P,j}C_2\bX_{13}
\end{matrix}\right]&=&\nonumber\\[0.2cm]
\left[\begin{matrix}
0&0&0\cr
-\tau\bX_{13}^T&-\tau\bX_{23}^T&-\tau\bX_{33}\cr
\bX_{12}^T+\bX_{13}^T&\bX_{22}+\bX_{23}^T&\bX_{23}+\bX_{33}
\end{matrix}\right]&&\nonumber\\[0.2cm]
+\mathbf{R}_jC_2\mathbf{X}_1&&
\end{eqnarray}
for
\begin{equation}\label{p7}
\bX_1=\begin{bmatrix}\bX_{11}&\bX_{12}&\bX_{13}\end{bmatrix}\in \mathbb{R}^{n_x\times (n_x+2n_u)}.
\end{equation}
Therefore, the bilinear constraints (\ref{wj}) are expressed by the linear constraints
\begin{eqnarray}\label{p5}
\mathbf{W}_j=\begin{bmatrix}
0&0&0\cr
-\tau\bX_{13}^T&-\tau\bX_{23}^T&-\tau\bX_{33}\cr
\bX_{12}^T+\bX_{13}^T&\bX_{22}+\bX_{23}^T&\bX_{23}+\bX_{33}
\end{bmatrix}+\mathbf{Y}_j,  \\ j=1, \dots, L \nonumber
\end{eqnarray}
plus the bilinear constraints
\begin{equation}\label{p6}
\mathbf{Y}_j=\mathbf{R}_jC_2\mathbf{X}_1, j=1, \dots. L.
\end{equation}
In other words, the BMI feasibility problem (\ref{bmib}) in $\mathbf{X}$, $\mathbf{R}$ and $\mathbf{W}$
is now equivalently transformed to the following BMI feasibility problem in  $\mathbf{X}$, $\mathbf{R}$, $\mathbf{W}$
and $\mathbf{Y}:=(\mathbf{Y}_1, \dots, \mathbf{Y}_L)$:
\begin{eqnarray}
(\ref{lct9b}), (\ref{c1a}), (\ref{c1b}), (\ref{p5}), (\ref{p6}),  \label{mob1}
\end{eqnarray}
where (\ref{lct9b}), (\ref{c1a}) and (\ref{c1b}) are linear matrix inequality (LMI) constraints, while (\ref{p5})
is linear constraints. The difficulty is now concentrated at $L$ bilinear constraints in (\ref{p6}),
in which only $\mathbf{X}_1$ is considered as a complicating variable that makes $L$ constraints in (\ref{p6})
nonlinear. Based on this observation, our strategy is to decouple this complicating variable $\mathbf{X}_1$ from (\ref{p6})
for a better treatment. Let us recall an auxiliary result.
\begin{mylem}\label{l-1}\cite{RTN14} For given matrix $W_{12},W_{22}$ of
sizes $n\times m$ and $m\times m$ with $W_{22}\succeq 0$, one has
\begin{equation}
\left(\begin{array}{ll}0&W_{12}\cr
W_{12}^T&W_{22}\end{array}\right)\succeq 0
\label{in4}
\end{equation}
if and only if $W_{12}=0$.
\end{mylem}
Using the above Lemma, we are now in a position to state the following result,
which is a cornerstone in handling bilinear constraints like
(\ref{p6}), which share a common complicating variable.
\begin{myth}\label{t-2} $L$ bilinear constraints in (\ref{p6}) are equivalently expressed by the following $L$ LMI constraints
\begin{eqnarray}
\left(\begin{array}{lll}\bW_{11,j}&\bY_j&\bR_j\cr
\bY_j^T&\bW_{22}&\bX_1^TC_2^T\cr
\bR_j^T&C_2\bX_1&I_{n_y}\end{array}\right)\succeq 0, \quad j=1,\dots, L,\label{in8}
\end{eqnarray}
plus the single bilinear constraint
\begin{equation}\label{sing}
\mathbf{W}_{22}=\bX_1^TC_2^TC_2\bX_1.
\end{equation}
\end{myth}
{\it Proof.} It can be easily seen that those $\mathbf{Y}_j$, $\mathbf{R}_j$ and $\mathbf{X}_1$ that are
constrained by (\ref{p6}) together with $\bW_{11,j}=\bR_j(\bR_j)^T$ and $\bW_{22}=C_2\bX_1(C_2\bX_1)^T$
are feasible for (\ref{in8}) and (\ref{sing}), showing the implication (\ref{p6})$\Rightarrow$(\ref{in8}) \& (\ref{sing}).

On the other hand, by Shur's complement, it follows from (\ref{in8}) that
\begin{eqnarray}
0&\preceq& \left(\begin{array}{lll}\bW_{11,j}&\bY_j\cr
\bY_j^T&\bW_{22}\end{array}\right)-
\left(\begin{array}{l}\bR_j\cr
\bX_1^TC_2^T\end{array}\right)\left(\begin{array}{ll}
\bR_j^T&C_2\bX_1\end{array}\right)\nonumber\\
&=&\left(\begin{array}{lll}\bW_{11,k}&\bY_k-\bR_kC_2\bX_1\cr
\bY_k^T-\bX_1^TC_2^T\bR_k^T&\mathbf{W}_{22}-\bX_1^TC_2^TC_2\bX_1\end{array}\right)\nonumber\\
&=&\left(\begin{array}{lll}\bW_{11,k}&\bY_k-\bR_kC_2\bX_1\cr
\bY_k^T-\bX_1^TC_2^T\bR_k^T&0\end{array}\right),\label{shur}
\end{eqnarray}
where we also used (\ref{sing}) in obtaining the last equality (\ref{shur}). Then applying Lemma \ref{l-1}
yields (\ref{p6}), showing the implication (\ref{in8}) \& (\ref{sing})$\Rightarrow$(\ref{p6}).\\ \qed

Now, the problem's nonconvexity is concentrated on the single constraint (\ref{sing}) that involves only $\mathbf{X}_1$.
\begin{myth}\label{auxt}
Under LMI constraints (\ref{in8}), the bilinear constraint (\ref{sing}) is equivalent to any from the two
following  constraints:\\
$(i)$ The matrix rank constraint
\begin{eqnarray}
\mbox{rank}(\clQ)=n_y\label{in11},
\end{eqnarray}
for
\begin{equation}\label{Q}
\clQ := \left( \begin{array}{ll} \bW_{22} & \bX_1^TC_2^T\cr C_2\bX_1 & I_{n_y} \end{array}\right);
\end{equation}
$(ii)$ The quadratic constraint
\begin{equation}\label{in14e}
\mbox{Trace}(\bW_{22})=||C_2\bX_1||^2
\end{equation}
\end{myth}
{\it Proof.} Note that (\ref{in8}) implies
\begin{equation}\label{qa}
\clQ\succeq 0
\end{equation}
which also yields
\begin{equation}\label{r1}
\mathbf{W}_{22} \succeq \bX_1^TC_2^TC_2\bX_1
\end{equation}
by Shur's complement. Also,
\[
\begin{array}{lll}
\mbox{rank}(\clQ)&=&\mbox{rank}(I_{n_y})+\mbox{rank}(\mathbf{W}_{22}-\bX_1^TC_2^TC_2\bX_1)\\
&=&n_y+\mbox{rank}(\mathbf{W}_{22}-\bX_1^TC_2^TC_2\bX_1),
\end{array}
\]
so (\ref{in11}) holds true if and only if $\mbox{rank}(\mathbf{W}_{22}-\bX_1^TC_2^TC_2\bX_1)=0$, which is
(\ref{sing}).\\
Next, it follows from (\ref{r1}) that (\ref{sing}) holds true if and only if
\[
\begin{array}{ll}
&\mbox{Trace}(\mathbf{W}_{22}-\bX_1^TC_2^TC_2\bX_1)=0\\
\Leftrightarrow&\mbox{Trace}(\mathbf{W}_{22})-||C_2\bX_1||^2=0\\
\Leftrightarrow&(\ref{in14e}).
\end{array}
\]
This completes the proof of Theorem \ref{auxt}.\\ \qed

The rank constraint (\ref{in11}) is discrete and absolutely intractable in general. However,
under condition (\ref{qa}), this rank constraint is
equivalent to the following continuous matrix-spectral constraint
\begin{equation}\label{cor2}
\mbox{Trace}(\clQ) - \lambda_{[n_y ]}(\clQ)=0,
\end{equation}
where $\lambda_{[n_y]}(\clQ)$ is the summation of the
$n_y$ largest eigenvalues of $\clQ$. Indeed, $\mbox{rank}(\clQ)\geq n_y$ but
(\ref{cor2}) means  $\clQ$ has at most $n_y$ nonzero eigenvalues so its rank is  $n_y$. \\
On the other hand, as
\[
\mbox{Trace}(\clQ) - \lambda_{[n_y ]}(\clQ)\geq 0,
\]
it follows from (\ref{cor2}) that
\begin{equation}\label{mes1}
\mbox{Trace}(\clQ) - \lambda_{[n_y ]}(\clQ)
\end{equation}
can be used to measure the degree of satisfaction
of the rank constraint (\ref{in11}). Instead of handling the nonconvex constraint (\ref{cor2}) we incorporate it into the objective, resulting in the following alternative formulation to (\ref{mob1})
\begin{subequations}\label{ub3}
\begin{eqnarray}
\ds\min_{\bX,\bW,\bR,\bY}\ F(\clQ):={\mbox{Trace}}(\clQ)-\lambda_{[n_y ]}(\clQ)\label{ub3a}\\
\quad\mbox{s.t.} \quad
(\ref{lct9b}),  (\ref{c1a}), (\ref{c1b}),  (\ref{p5}), (\ref{in8}).\label{ub3b}
\end{eqnarray}
\end{subequations}
Suppose $X_1^{(\kappa)}$ and $W_{22}^{(\kappa)}$ are feasible for (\ref{ub3}). Set
\[
\clQ^{(\kappa)} := \left( \begin{array}{ll} W_{22}^{(\kappa)} &
(X_1^{(\kappa)})^TC_2^T\cr C_2X_1^{(\kappa)} & I_{n_y} \end{array}\right)
\]
Function $\lambda_{[n_y]}(\clQ)$ is nonsmooth but is lower bounded by the linear function
\begin{equation}\label{lm1}
\sum_{i=1}^{n_y}(w_{i}^{(\kappa)})^T\clQ w_{i}^{(\kappa)},
\end{equation}
where $w_{1}^{(\kappa)},....,w_{n_y}^{(\kappa)}$ are the normalized eigenvectors corresponding to
$n_y$ largest eigenvalues of $\clQ^{(\kappa)}$.
Thus, the following convex optimization problem provides an upper bound for the nonconvex optimization
problem (\ref{ub3}),
\begin{equation}\label{ub4a}
\begin{array}{l}
\ds \min_{\bX,\bW,\bR,\bY}\ F^{(\kappa)}(\clQ): = {\mbox{Trace}}(\clQ)-\sum_{i=1}^{n_y}(w_{i}^{(\kappa)})^T \clQ w_{i}^{(\kappa)} \\
\quad\mbox{s.t.} \quad (\ref{ub3b}).
\end{array}
\end{equation}
Suppose that $(X_1^{(\kappa+1)},W_{22}^{(\kappa+1)})$ is the optimal solution of
(\ref{ub4a}) and
\[
\clQ^{(\kappa+1)} := \left( \begin{array}{ll} W_{22}^{(\kappa+1)} &
(X_1^{(\kappa+1)})^TC_2^T\cr C_2X_1^{(\kappa+1)} & I_{n_y} \end{array}\right).
\]
Then
\[
\begin{array}{lll}
F(\clQ^{(\kappa+1)})&\leq &F^{(\kappa)}(\clQ^{(\kappa+1)})\\
&<&F^{(\kappa)}(\clQ^{(\kappa)})\\
&=&F(\clQ^{(\kappa)}),
\end{array}
\]
as far as $\clQ^{(\kappa+1)}\neq \clQ^{(\kappa)}$,
implying that $\clQ^{(\kappa+1)}$ is better than $\clQ^{(\kappa)}$
towards optimizing (\ref{ub4a}). Similarly to \cite{PTKN12}, we establish the following result.
\begin{mypro}\label{conpro}
Initialized by any feasible point $\clQ^{(0)}$ for the
convex constraints (\ref{ub3b}), $\{\clQ^{(\kappa)}\}$ is a sequence of improved
feasible points of the nonconvex optimization problem (\ref{ub3}), which converges to a point satisfying
the first-order necessary optimality conditions.
\end{mypro}

In  Algorithm \ref{alg1} we propose a convex programming based computational procedure
for the nonconvex optimization problem (\ref{ub3}).

\begin{algorithm}[!t]\caption{Nonconvex Spectral Optimization Algorithm for Solving BMI feasibility}\label{alg1}
  \begin{algorithmic}[1]
  \State {\bf Initialization.} Set $\kappa := 0$ and solve the LMI (\ref{ub3b})
to find a feasible point $(X^{(\kappa)},W^{(\kappa)},R^{(\kappa)},Y^{(\kappa)})$.
Given computational tolerance $\epsilon>0$, stop the algorithm  and
accept $(X^{(0)},W^{(0)},R^{(0)},Y^{(0)})$ as the solution of BMI (\ref{bmib}) if
\begin{equation}
F(\clQ^{(\kappa)}) \leq \epsilon.
\end{equation}
  \Repeat
  \State Solve the convex optimization problem (\ref{ub4a}), to find the optimal solution $(X^{(\kappa+1)},W^{(\kappa+1)},R^{(\kappa+1)},Y^{(\kappa+1)})$
  \State Set $\kappa := \kappa+1$.
  \Until{\begin{equation}\label{stopc}
\frac{F(\clQ^{(\kappa-1)})-F(\clQ^{(\kappa)})}{F(\clQ^{(\kappa-1)})} \leq \epsilon.
\end{equation} }
   \State Accept $(X^{(\kappa)},W^{(\kappa)},R^{(\kappa)},Y^{(\kappa)})$ as the solution of (\ref{ub3}).
Accept  $(X^{(\kappa)},W^{(\kappa)},R^{(\kappa)},Y^{(\kappa)})$ as the solution of BMI (\ref{bmib})
if $F(\clQ^{(\kappa)})\leq \epsilon$. Otherwise declare that BMI (\ref{bmib}) is infeasible.
  \end{algorithmic}
\end{algorithm}

So far, in solving (\ref{ub3})  we are based on (\ref{mes1}) as
the satisfaction degree of the rank constraint (\ref{in11}) and thus of the bilinear constraint (\ref{sing}).
For larger value of $n_y$,  Algorithm \ref{alg1} may converge slowly. We now use
\begin{equation}\label{mes2}
1-\frac{||C_2\bX_1||^2}{\mbox{Trace}(\bW_{22})}
\end{equation}
as an alternative degree for satisfaction of the bilinear constraint (\ref{sing}) because according to  (\ref{r1}),
(\ref{mes2}) is positive and by (\ref{in14e}), it is zero if and only if the bilinear constraint (\ref{sing}) is
satisfied. Accordingly, instead of  (\ref{ub3})  we use the following optimization problem:
\begin{eqnarray}\label{al1}
\ds\min_{\bX,\bW,\bR,\bY}\ -\frac{||C_2\bX_1||^2}{\mbox{Trace}(\bW_{22})}
\quad\mbox{s.t.} \quad
(\ref{ub3b}).
\end{eqnarray}
Note that function $g(\bX_1,\bW_{22}):=||C_2\bX_1||^2/\mbox{Trace}(\bW_{22})$ is convex in $\bX_1$ and $\bW_{22}\succeq 0$ \cite{Tuybook},
so
\[
\begin{array}{lll}
g(\bX_1,\bW_{22})&\geq&g(X_1^{(\kappa)},W_{22}^{(\kappa)})+\la \nabla g(X_1^{(\kappa)},W_{22}^{(\kappa)}),
\\&&(\bX_1,\bW_{22})-(X_1^{(\kappa)},W_{22}^{(\kappa)})\ra\\[0.3cm]
&=&-2\ds\frac{ \mbox{Trace}((X_1^{(\kappa)})^TC_2^TC_2\bX_1)}{\mbox{Trace}(W_{22}^{(\kappa)})}
\\&&-\ds\frac{||C_2X_1^{(\kappa)}||^2 \mbox{Trace}(\bW_{22})}{(\mbox{Trace}(W_{22}^{(\kappa)}))^2}.
\end{array}
\]
Thus, instead of (\ref{ub4a}), we solve the following convex optimization problem, which is an upper bound for
the nonconvex optimization problem (\ref{al1}), to generate $(X^{(\kappa+1)},W^{(\kappa+1)},R^{(\kappa+1)},Y^{(\kappa+1)})$
at the $\kappa$-th iteration:
\begin{equation}\label{th3}
\begin{array}{ll}
\ds \min_{\bX,\bW,\bR,\bY}\ & -2\ds\frac{ \mbox{Trace}((X_1^{(\kappa)})^TC_2^TC_2\bX_1)}{\mbox{Trace}(W_{22}^{(\kappa)})}\\
&+\ds\frac{||C_2X_1^{(\kappa)}||^2 \mbox{Trace}(\bW_{22})}{(\mbox{Trace}(W_{22}^{(\kappa)}))^2}\quad\\
\quad\mbox{s.t.}&\quad (\ref{ub3b}).
\end{array}
\end{equation}
A pseudo-code for the computational procedure, which is based on computation for (\ref{th3}) at each iteration, is described
by Algorithm \ref{alg2}.
\begin{algorithm}[!t]\caption{Fractional Optimization Algorithm for Solving BMI feasibility}\label{alg2}
  \begin{algorithmic}[1]
  \State {\bf Initialization.} Set $\kappa := 0$ and solve the LMI (\ref{ub3b})
to find a feasible point $(X^{(\kappa)},W^{(\kappa)},R^{(\kappa)},Y^{(\kappa)})$.
Given computational tolerance $\epsilon>0$, stop the algorithm  and
accept $(X^{(0)},W^{(0)},R^{(0)},Y^{(0)})$ as the solution of BMI (\ref{bmib}) if
\begin{equation}
1-g(X_1^{(\kappa)}, W_{22}^{(\kappa)}) \leq \epsilon.
\end{equation}
  \Repeat
  \State Solve the convex optimization problem (\ref{th3}), to find the optimal solution $(X^{(\kappa+1)},W^{(\kappa+1)},R^{(\kappa+1)},Y^{(\kappa+1)})$
  \State Set $\kappa := \kappa+1$.
  \Until{\begin{equation}\label{stopc2}
\frac{g(X_1^{(\kappa)}, W_{22}^{(\kappa)})-g(X_1^{(\kappa-1)}, W_{22}^{(\kappa-1)})}{g(X_1^{(\kappa-1)}, W_{22}^{(\kappa-1)})} \leq \epsilon.
\end{equation} }
   \State Accept $(X^{(\kappa)},W^{(\kappa)},R^{(\kappa)},Y^{(\kappa)})$ as the solution of (\ref{ub3}).
Accept  $(X^{(\kappa)},W^{(\kappa)},R^{(\kappa)},Y^{(\kappa)})$ as the solution of BMI (\ref{bmib})
if $1-g(X_1^{(\kappa)}, W_{22}^{(\kappa)})\geq \epsilon$. Otherwise declare that BMI (\ref{bmib}) is infeasible.
  \end{algorithmic}
\end{algorithm}
\section{Simulation results}
An important step is to check if there is a controller (\ref{if2}) to stabilize system (\ref{ts1}).
Define the  block $(1,1)$ in (\ref{PLMI2}) as
\[
\widetilde{\clM}_{ij}(\mathbf{X},\mathbf{R}_j,\mathbf{W}_j):=(A_{0i}\mathbf{X}+\clB_i\mathbf{W}_j)+(*).
\]
Then the existing of a stabilizing controller (\ref{if2}) is guaranteed by the feasibility of the system consisting
of (\ref{lct9b}), (\ref{wj}) and
\begin{eqnarray}
\widetilde{\clM}_{ii}(\mathbf{X},\mathbf{R}_i,\mathbf{W}_i)\prec 0, \quad i=1,\dots, L,\label{new1}\\
\ds\frac{1}{L-1}\widetilde{\clM}_{ii}(\mathbf{X},\mathbf{R}_i,\mathbf{W}_i)
+\frac{1}{2}(\widetilde{\clM}_{ij}(\mathbf{X},\mathbf{R}_j,\mathbf{W}_j)+ \nonumber \\
\widetilde{\clM}_{ji}(\mathbf{X},\mathbf{R}_i,\mathbf{W}_i))\prec 0,\label{new2} \quad 1\leq i\neq j\leq L.
\end{eqnarray}
Thus, we can use Algorithm \ref{alg1} or Algorithm \ref{alg2} to check its feasibility, which invokes either the
convex optimization problem
\begin{equation}\label{new3}
\begin{array}{l}
\ds \min_{\bX,\bW,\bR,\bY}\ {\mbox{Trace}}(\clQ)-\sum_{i=1}^{n_y}(w_{i}^{(\kappa)})^T \clQ w_{i}^{(\kappa)}\\
\quad\mbox{s.t.} \quad (\ref{lct9b}), (\ref{wj}), (\ref{new1}), (\ref{new2}),
\end{array}
\end{equation}
or the convex optimization problem
\begin{equation}\label{new4}
\begin{array}{ll}
\ds \min_{\bX,\bW,\bR,\bY}\ & -2\ds\frac{ \mbox{Trace}((X_1^{(\kappa)})^TC_2^TC_2\bX_1)}{\mbox{Trace}(W_{22}^{(\kappa)})}\\
&+\ds\frac{||C_2X_1^{(\kappa)}||^2 \mbox{Trace}(\bW_{22})}{(\mbox{Trace}(W_{22}^{(\kappa)}))^2}\quad\\
\\ \quad\mbox{s.t.}&\quad (\ref{lct9b}), (\ref{wj}), (\ref{new1}), (\ref{new2})
\end{array}
\end{equation}
instead of (\ref{ub4a}) or (\ref{th3}) at the $\kappa$-th iteration to generate the next iterative point $(X^{(\kappa+1)}, W^{(\kappa+1)}, R^{(\kappa+1)}, Y^{(\kappa+1)})$.\\
Whenever, a feasible point $(X^{(\kappa)}, W^{(\kappa)}, R^{(\kappa)}, Y^{(\kappa)})$ of (\ref{lct9b}), (\ref{wj}), (\ref{new1})
and (\ref{new2}) is found, we solve the following convex optimization problem to determine the initial $\gamma_u$ for the bisection procedure:
\[
\begin{array}{rll}
\ds\min_{\gamma} \gamma\quad\mbox{s.t.}\quad \clM_{ii}(X^{(\kappa)},R_i^{(\kappa)},W_i^{(\kappa)},\gamma)&\prec& 0,\\
\ds\frac{1}{L-1}\clM_{ii}(X^{(\kappa)},R_i^{(\kappa)},W_i^{(\kappa)},\gamma)&&\\
+\ds\frac{1}{2}(\clM_{ij}(X^{(\kappa)},R_j^{(\kappa)},W_j^{(\kappa)},\gamma)&&\\
+\ds\clM_{ji}(X^{(\kappa)},R_i^{(\kappa)},W_i^{(\kappa)},\gamma))&\prec& 0,\\
1\leq i\neq j\leq L. &&
\end{array}
\]
\subsection{Inverted pendulum control}
\begin{figure}[h]
\centering
\includegraphics[width=0.9 \columnwidth]{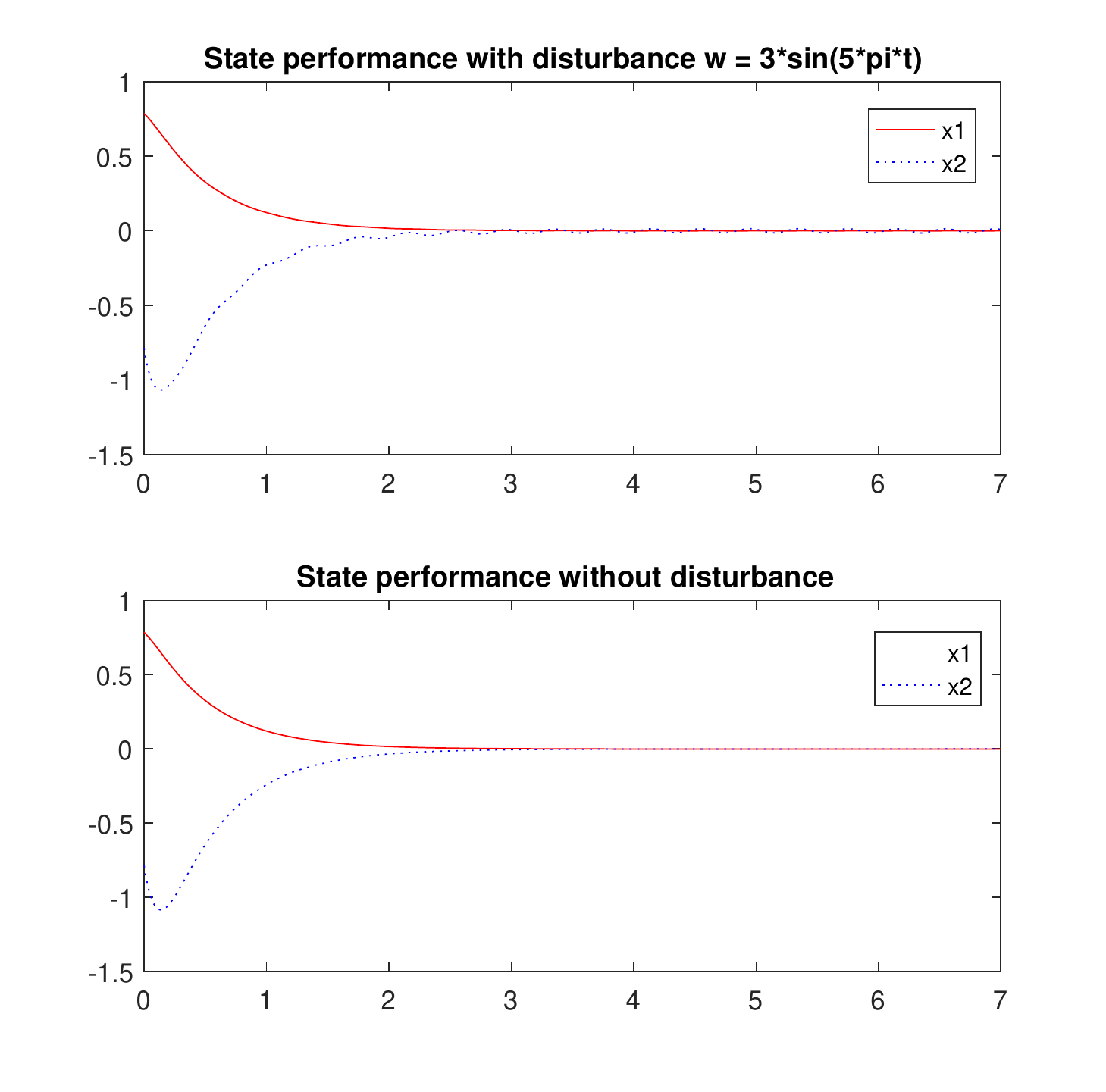}
\caption{The state behaviour  with and without disturbance}
\label{state_KC2}
\end{figure}

\begin{figure}[h]
\centering
\includegraphics[width=0.9 \columnwidth]{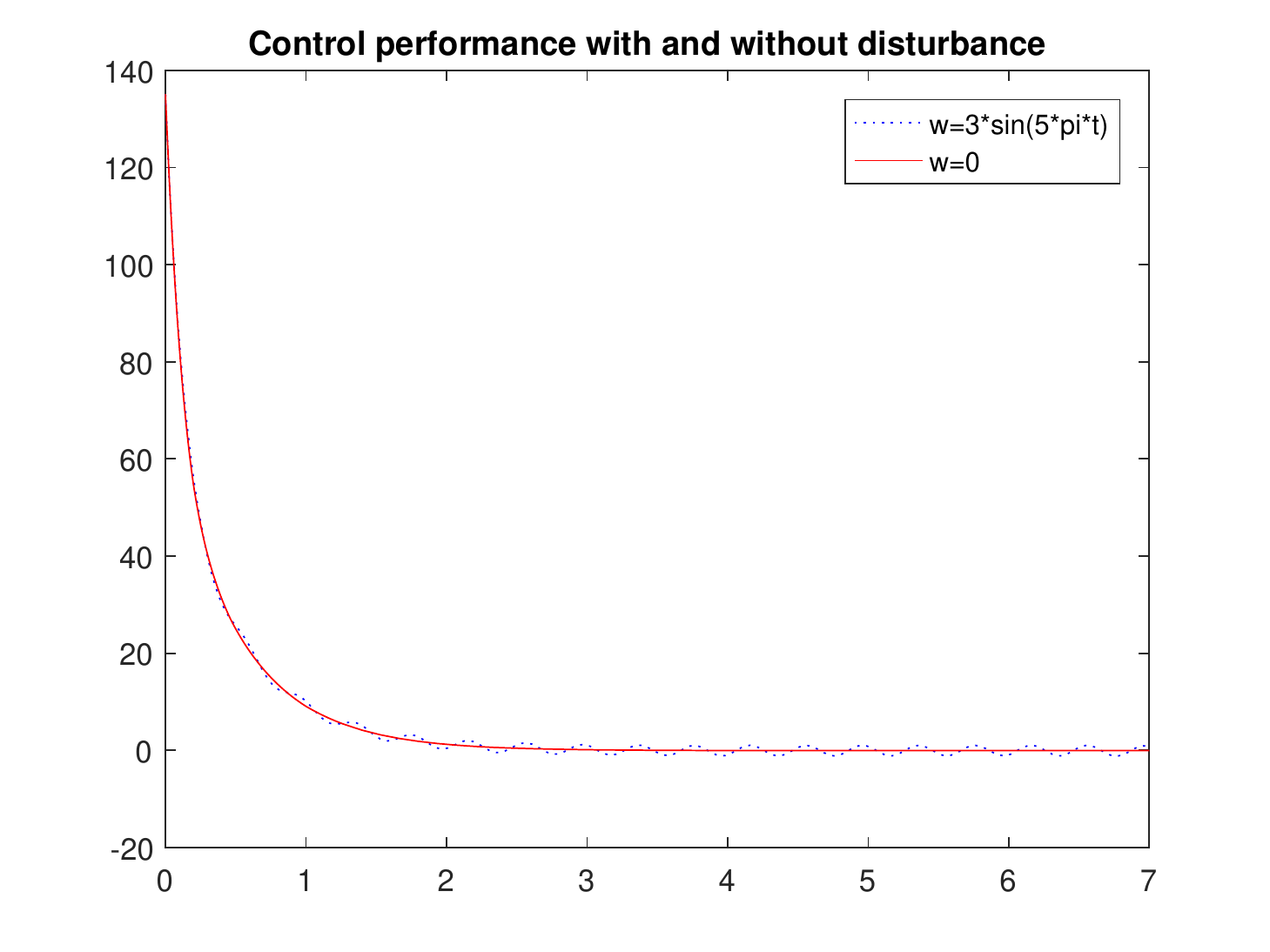}
\caption{PID PDC behaviour with and without the disturbance }
\label{control_KC2}
\end{figure}

The motion of an inverted pendulum system with a point mass of mass $m=2$ kg, a rigid rod of the length $\ell=0.5m$ and a cart
of mass $M=8$ kg can be described by (\ref{ts2}) \cite{CRF96} with $L = 2$ and
\[
\begin{array}{c}
A_1 =
\begin{bmatrix}
0 & 1\\
17.2941 & 0
\end{bmatrix},\quad
A_2 =
\begin{bmatrix}
0 & 1\\
12.6305 & 0
\end{bmatrix},\\
B_{11} = B_{12} =
\begin{bmatrix}
0 \\
{\color{black} 0.1}
\end{bmatrix},\quad
B_{21} =
\begin{bmatrix}
0 \\
-0.1765
\end{bmatrix},\\
B_{22} =
\begin{bmatrix}
0 \\
-0.0779
\end{bmatrix},\quad
C_{11} = C_{12} =
\begin{bmatrix}
1 & 1
\end{bmatrix},\quad
C_2 = \begin{bmatrix}
3 & 0
\end{bmatrix},\\
D_{11,i} \equiv 0.1, \quad
D_{12,i} \equiv 0, \quad
D_{21} = 0.
\end{array}
\]
The system state is $x=(x_1, x_2)^T$, where $x_1$ is the angle measured from the inverted equilibrium position
(angular position) and $x_2$ is the angular velocity. The membership functions in (\ref{alpha1}) are
\begin{equation}
\begin{array}{ll}
\alpha_1(t) =& (1-(1+e^{(-7(x_1(t)-\pi/4))})^{-1})\cdot \\
&(1+e^{(-7(x_1(t)+\pi/4))})^{-1}\\
\alpha_2(t) =& 1 - \alpha_1(t), \quad x_1(t) \in [-\pi/3,\pi/3].
\end{array}
\end{equation}
Based on the measured output $y=x_1(t)$ the task of the PID control is to minimize the effect of the disturbance
in stabilizing the system. Therefore, the controlled output is set as $z=x_1+x_2$.\\
In this example, $\tau = 6$ is set for (\ref{if2}).
The minimal $\gamma=0.12$ is obtained by using the bisection procedure.
At $\gamma=0.12$, Algorithm \ref{alg1} needs $4$ iterations to arrive the following numerical values
for implementing  PID PDC (\ref{if2}):
$R_{P1} = 72.3777$, $R_{P2} = 99.2379$, $R_{I1} = 0.1449$, $R_{I2} = 0.1028$, $R_{D1} =  5.0864$ and $R_{D2} = 8.8573$.
Figs. \ref{state_KC2}-\ref{control_KC2} respectively
show the behavior of the system state and control  with disturbance $w=3\sin(5\pi t)$
and  with no disturbance.
The initial state is $x(0) = (\pi/4,-\pi/4)^T$.
The obtained PID PDC stabilizes the inverted pendulum system well in the both scenarios. The system state motion
and control load are very smooth compared with \cite[Fig. 2]{CGLV16}

\begin{figure}[h]
\centering
\includegraphics[width=0.8 \columnwidth]{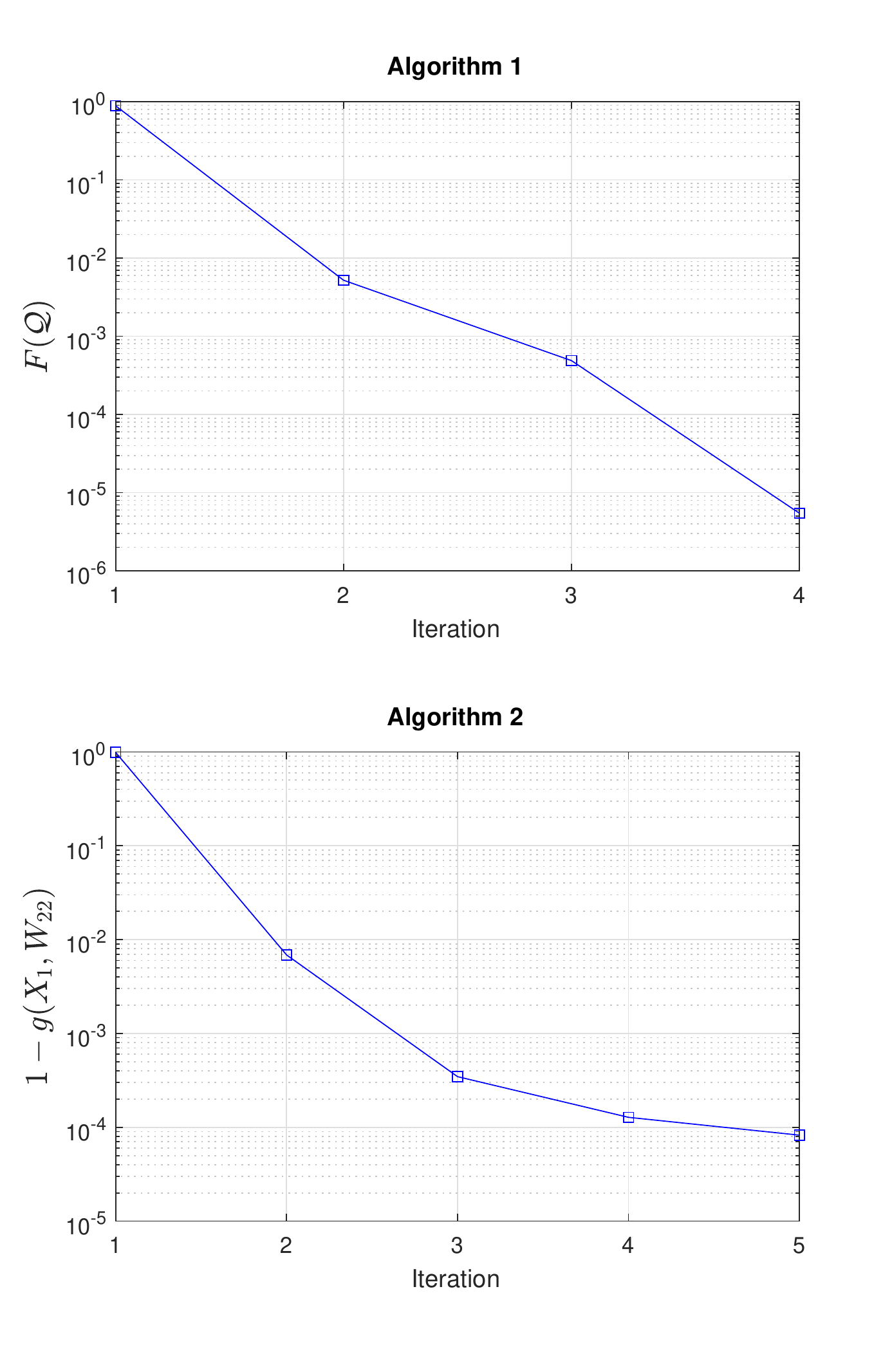}
\caption{Convergence performance by Algorithm \ref{alg1} and Algorithm \ref{alg2} for the inverted pendulum system}
\label{caseA_Alg1_Alg2_iteration}
\end{figure}

Algorithm \ref{alg2} achieves worse
$\gamma=0.13$ and needs $5$ iterations for convergence for $\gamma=0.13$.
Fig.\ref{caseA_Alg1_Alg2_iteration} show the convergence behaviour of Algorithm \ref{alg1}
(for $\gamma=0.12$) and Algorithm \ref{alg2} (for $\gamma=0.13$).

\subsection{Duffing forced-oscillation}
\begin{figure}[h]
\centering
\includegraphics[width=0.8 \columnwidth]{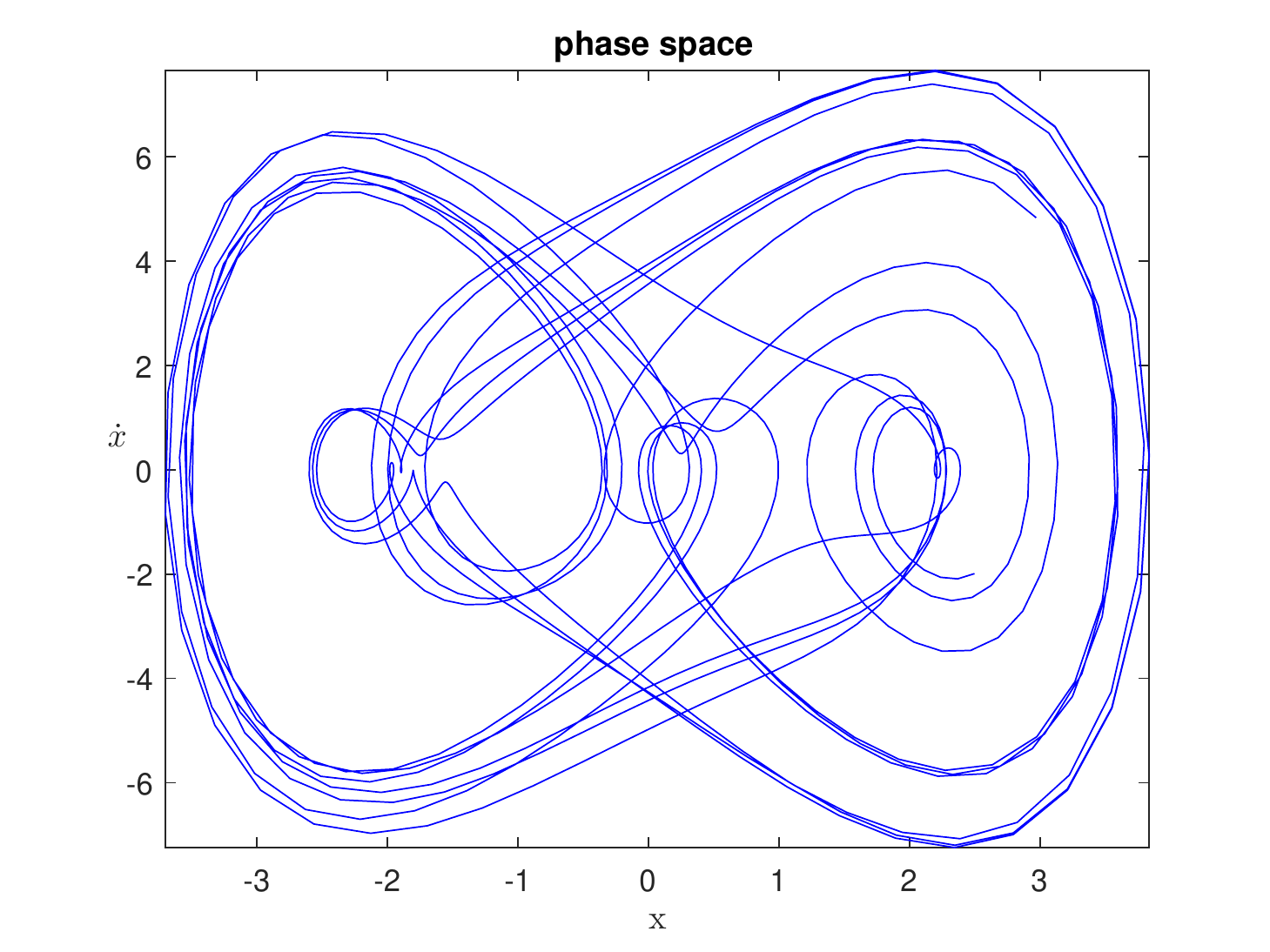}
\caption{The system state behavior  without control}
\label{caseB_chaotic}
\end{figure}

By \cite{TW04}, the Duffing forced-oscillation equation
\[
\ddot{\mathbf{x}}+0.2\dot{\mathbf{x}}+\mathbf{x}^3-10\cos t -u(t)=0
\]
with control input $u(t)$ and measured output $\mathbf{x}$ can be described by  (\ref{ts2}) with $L = 2$ and
\[
\begin{array}{c}
A_1 =
\begin{bmatrix}
0 & 1\\
0 & -0.2
\end{bmatrix},\quad
A_2 =
\begin{bmatrix}
0 & 1\\
-d^2 & -0.2
\end{bmatrix},\\
B_{11} = B_{12} =
\begin{bmatrix}
0 \\
0.1
\end{bmatrix},\quad
B_{21} = B_{22} =
\begin{bmatrix}
0 \\
1
\end{bmatrix},\\
C_{11} = C_{12} =
\begin{bmatrix}
1 & 1
\end{bmatrix},\quad
C_2 = \begin{bmatrix}
1 & 0
\end{bmatrix},\\
D_{11,i} \equiv 0.1, \quad  D_{12,i} \equiv 0, \quad
D_{21} = 0.
\end{array}
\]
The membership functions in (\ref{alpha1}) are
\[
\alpha_1(t) = 1-\frac{x^2_1(t)}{d^2}, \alpha_2(t) = \frac{x_1^2(t)}{d^2},  x_1(t) \in [-d,d].
\]
The system state is $x=(\mathbf{x},\dot{\mathbf{x}})^T$  but only $\mathbf{x}$ is measurable so
$y=\mathbf{x}$. The task is to minimize the effect of the disturbance $w(t)$ in stabilizing the system, so
the controlled output is set as $z=\mathbf{x}+\dot{\mathbf{x}}$.
The reader is also referred to \cite[IV.B]{TANK04} for a different form of fuzzy systems for this oscillation.
Without the control input $u(t)$ the system state behaviour is chaotic as Fig. \ref{caseB_chaotic} shows.
Since $x_1$ is always in the region $[-4,4]$ we can set $d=4$.

We set $\tau = 2$ for (\ref{if2}) in this example.
The minimal $\gamma=1.1$ is obtained by the bisection procedure. For this value of $\gamma$, Algorithm \ref{alg1} need $10$ iterations to arrive the following numerical values for implementing
PID PDC (\ref{if2}): $R_{P1} =  -96.8448$, $R_{P2} = 6.4360$, $R_{I1} =-1.4964$, $R_{I2}= -1.4984$, $R_{D1} =  -0.7271$ and $R_{D2} = -0.0094$.
%\footnote{\color{red}These high gains are silly. The reviewers will reject the paper immediately
%to see these silly results}

\begin{figure}[h]
\centering
\includegraphics[width=0.8 \columnwidth]{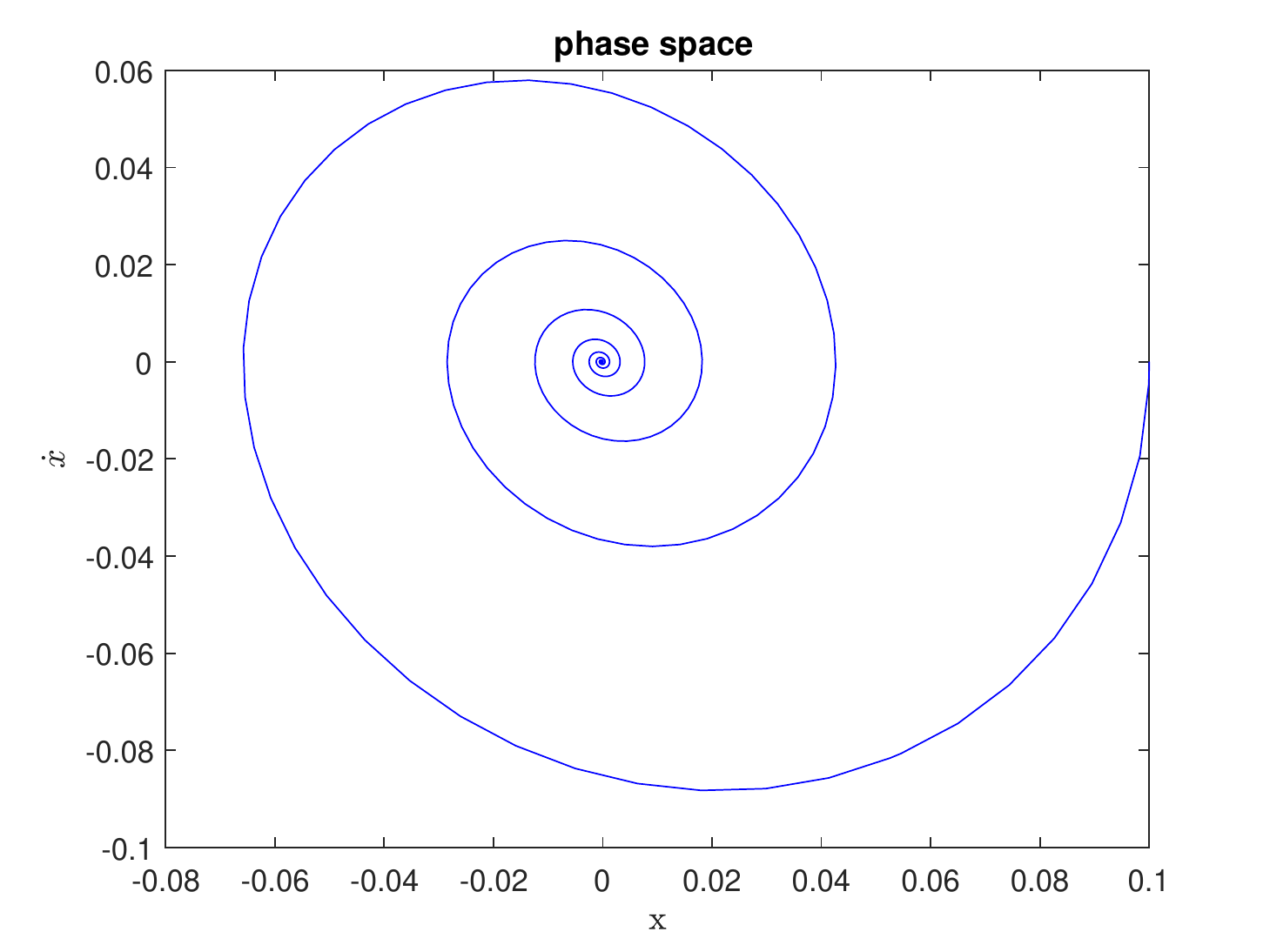}
\caption{The system state under PID PDC control}
\label{caseB_phase}
\end{figure}

\begin{figure}[h]
\centering
\includegraphics[width=0.9 \columnwidth]{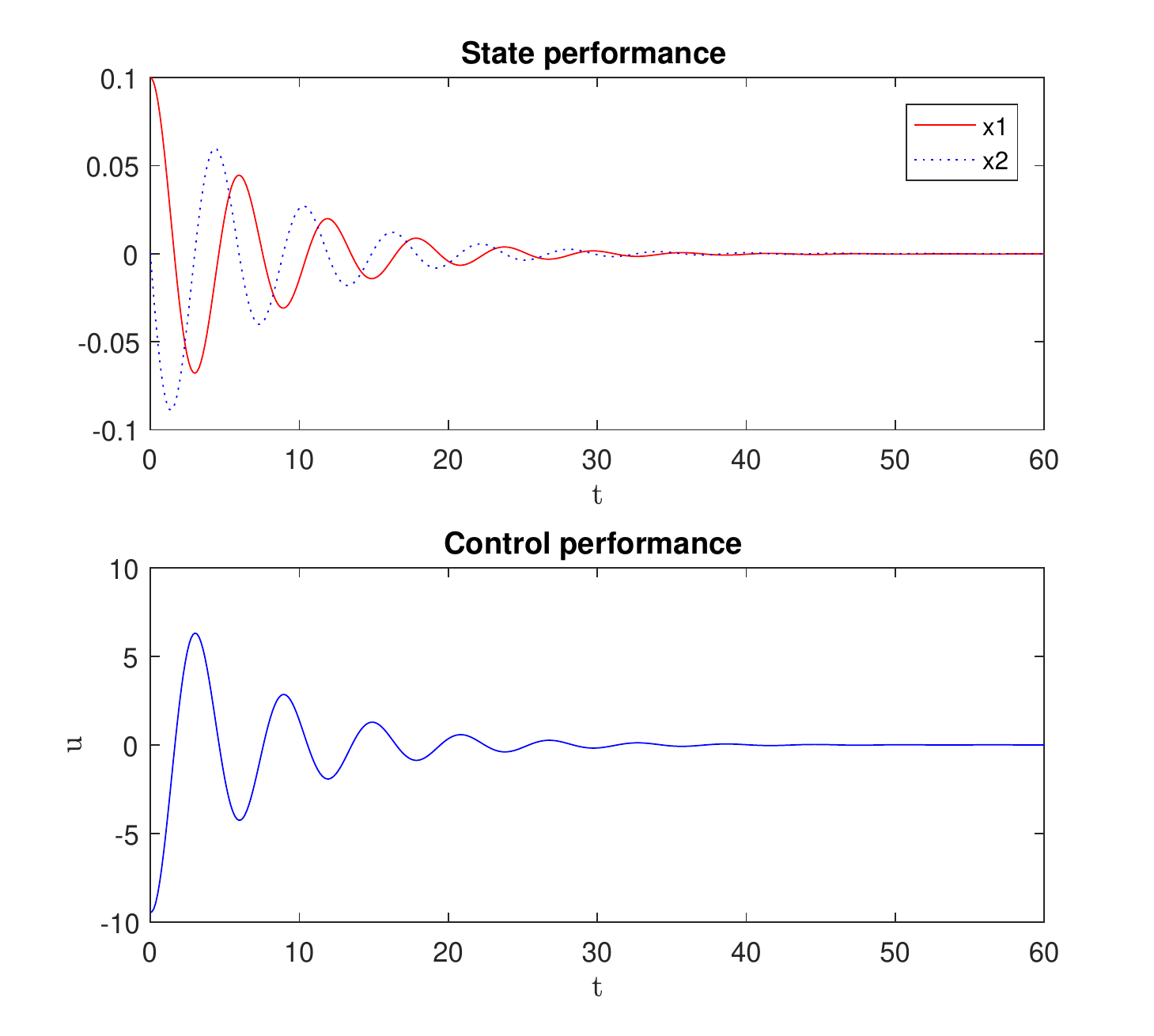}
\caption{The state and PID PDC behavior of the Duffing forced-oscillation system}
\label{caseB_state_control}
\end{figure}

Fig. \ref{caseB_phase} represents the state plane with PID PDC. The initial state condition $x(0) = (0.1,0)^T$.
Fig. \ref{caseB_state_control} depicts the behavior of
the state and PID PDC.
%\footnote{\color{red}Another
%silly mistake. In Duffing equation, $\ddot{x}+\delta \dot{x}+\alpha x+\beta x^3+\gamma \cos(\omega t)=0$
%the disturbance is always $\gamma\cos(\omega t)$. You don't understand any thing. You simulate a quite different system
%and you even don't know that the disturbance is already included in the equation
%}
Again the PID PDC stabilizes the Duffing forced-oscillation system well.
%\footnote{\color{black}You follow \cite{CGLV16}, which
%is a bad paper in general. You should provide Figures like \cite[Fig.6]{TANK04}}

\begin{figure}[h]
\centering
\includegraphics[width=0.8 \columnwidth]{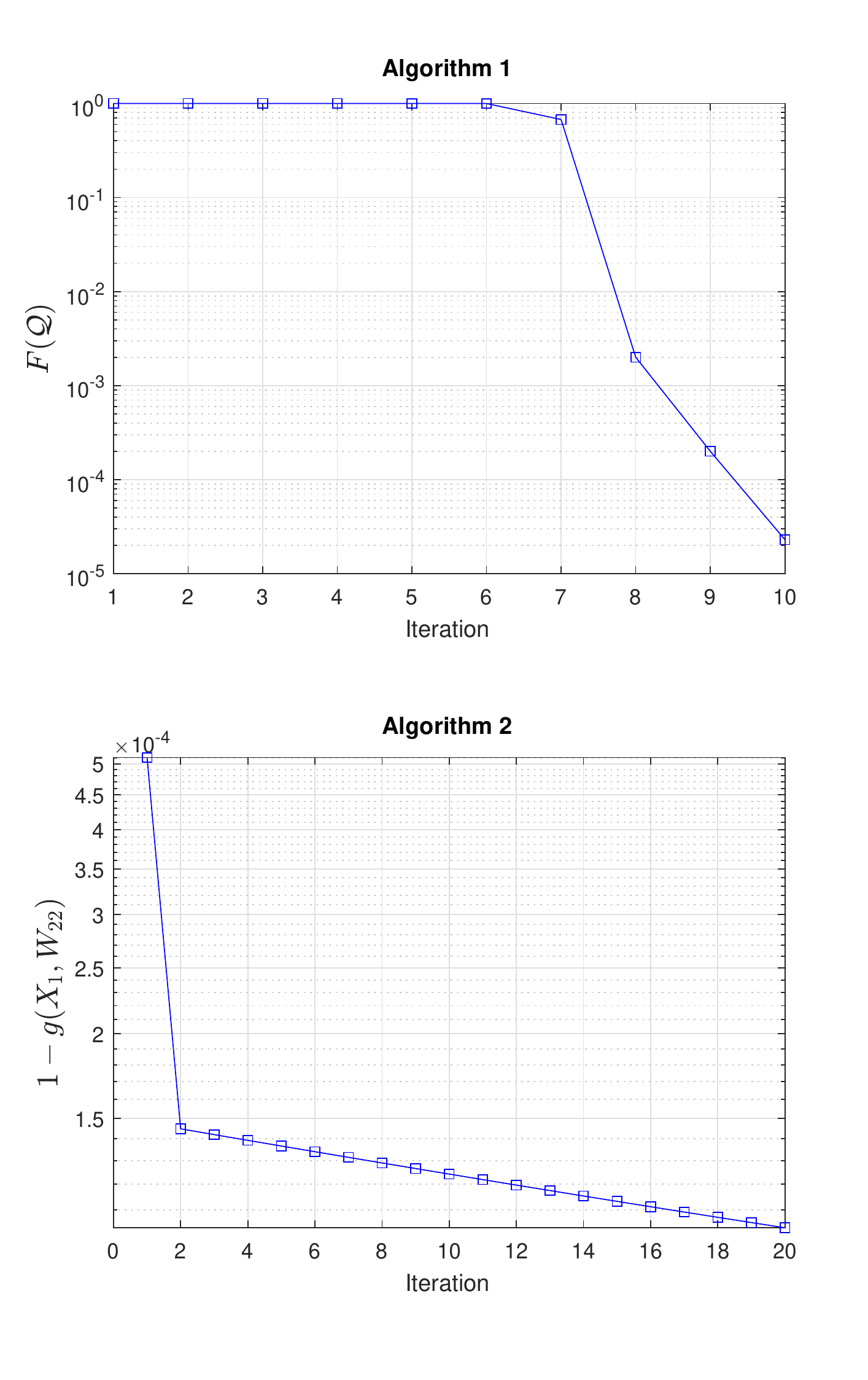}
\caption{Convergence performance by Algorithm \ref{alg1} and Algorithm \ref{alg2} for the Duffing forced-oscillation system}
\label{caseB_Alg1_Alg2_iteration}
\end{figure}

Meanwhile Algorithm \ref{alg2} achieves worse $\gamma=1.4$ and needs $20$ iterations for converge for this value
of $\gamma$. Fig.\ref{caseB_Alg1_Alg2_iteration} shows the convergence behaviour of  Algorithm \ref{alg1}
 (for $\gamma=1.1$) and Algorithm \ref{alg2} (for $\gamma=1.4$).
  Their convergence is dependent on initial points. Algorithm \ref{alg1} converges not rapidly until the seventh iteration, while Algorithm \ref{alg1} converges rapidly after the first iteration.

\subsection{TORA}
By \cite{TIW98} and \cite{TANY01},  the eccentric rotational proof mass actuator (TORA) system
can be represented by T-S model (\ref{ts2}) with
\[
L = 4, \quad \alpha = 0.99, \quad \phi = 0.1, \quad c = 4,
\]
\[
B_{1i} \equiv 0, \quad D_{11i} \equiv 0, \quad D_{12i} \equiv 0, \quad D_{21i} \equiv 0,
\]
\[A_1 =
\begin{bmatrix}
0 & 1 & 0 & 0\\
-1 & 0 & \epsilon\sin(\alpha\pi)/(\alpha\pi) & 0\\
0 & 0 & 0 & 1\\
-\phi/(1-\phi^2)& 0 & 0 & 0
\end{bmatrix},
\]
\[
A_2 =
\begin{bmatrix}
0 & 1 & 0 & 0\\
-1 & 0 & 2\phi/\pi & 0\\
0 & 0 & 0 & 0\\
0& 0 & 0 & 0
\end{bmatrix},
\]
\[
A_3 =
\begin{bmatrix}
0 & 1 & 0 & 0\\
-1 & 0 & \phi & 0\\
0 & 0 & 0 & 1\\
\phi/(1-\phi^2)& 0 & -\phi^2/(1-\phi^2) & 0
\end{bmatrix},
\]
\[
A_4 =
\begin{bmatrix}
0 & 1 & 0 & 0\\
-1 & 0 & \phi & 0\\
0 & 0 & 0 & 1\\
\phi/(1-\phi^2)& 0 & -\phi^2(1-c^2)/(1-\phi^2) & 0
\end{bmatrix},
\]
\[
B_{21} =
\begin{bmatrix}
0 \\
0 \\
0 \\
1/(1-\phi^2)
\end{bmatrix},
B_{22} =
\begin{bmatrix}
0 \\
0 \\
0 \\
1
\end{bmatrix},
\]
\[
B_{23} =
\begin{bmatrix}
0 \\
0 \\
0 \\
1/(1-\phi^2)
\end{bmatrix},
B_{24} =
\begin{bmatrix}
0 \\
0 \\
0 \\
1/(1-\phi^2)
\end{bmatrix},
\]
\[
C_{1i}
\begin{bmatrix}
1 & 0 & 0 & 0\\
0 & 0 & 1 & 0
\end{bmatrix},
C_2 = \begin{bmatrix}
1 & 0 & 1 & 0\\
0 & 1 & 0 & 1\\
1 & 1 & 0 & 0
\end{bmatrix},
\]
The membership functions in (\ref{alpha1}) are
\[
\begin{array}{c}
\alpha_1(t) = \frac{x_1^2(t)}{a^2},\quad
\alpha_2(t) = \frac{1}{2}-\alpha_1(t),\\
\alpha_3(t) = \frac{b\sin(x_3(t))-x_3(t)\sin(b)}{x_3(t)(b-\sin(b))},\quad
\alpha_4(t) = \frac{1}{2}-\alpha_3(t),
\end{array}
\]
with $a = 0.8$, $ b = 0.6 $, and $x_1(t) \in [-a,a]$ and $x_3(t) \in [-b,b]$. The system state is $x=(x_1,x_2,x_3,x_4)$, where $x_3=\theta$ and $x_4=\dot{\theta}$ are the angular position
and angular velocity of the rotational proof mass, and $x_1=\bar{x}_1+\epsilon \sin x_3$, $x_2=\bar{x}_2+\epsilon x_4\cos x_3$
with $\bar{x}_1=q$ and $\bar{x}_2=\dot{q}$ the translational position and velocity of the cart. In this application, only the translation position and angular position are measurable so $y=(x_1, x_3)^T$. The main task is to
minimize the effect of the disturbance $w$ in regulating the translation and angular positions to the equilibrium so
the controlled output is set as  $z=(x_1,x_3)^T$.

\begin{figure}[h]
\centering
\includegraphics[width=0.9 \columnwidth]{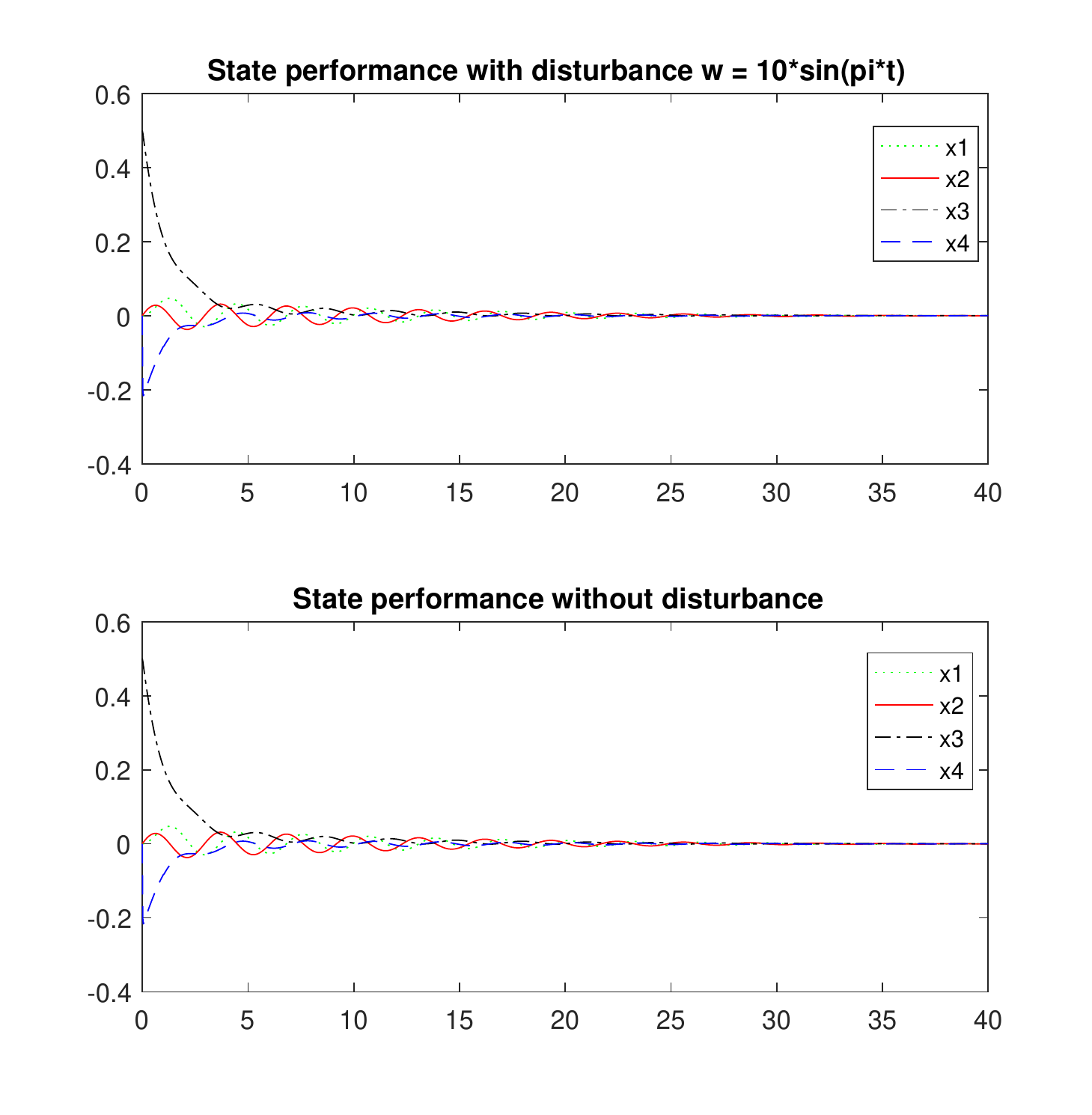}
\caption{The state behaviour with and without disturbance}
\label{state_TORA}
\end{figure}

\begin{figure}[h]
\centering
\includegraphics[width=0.8 \columnwidth]{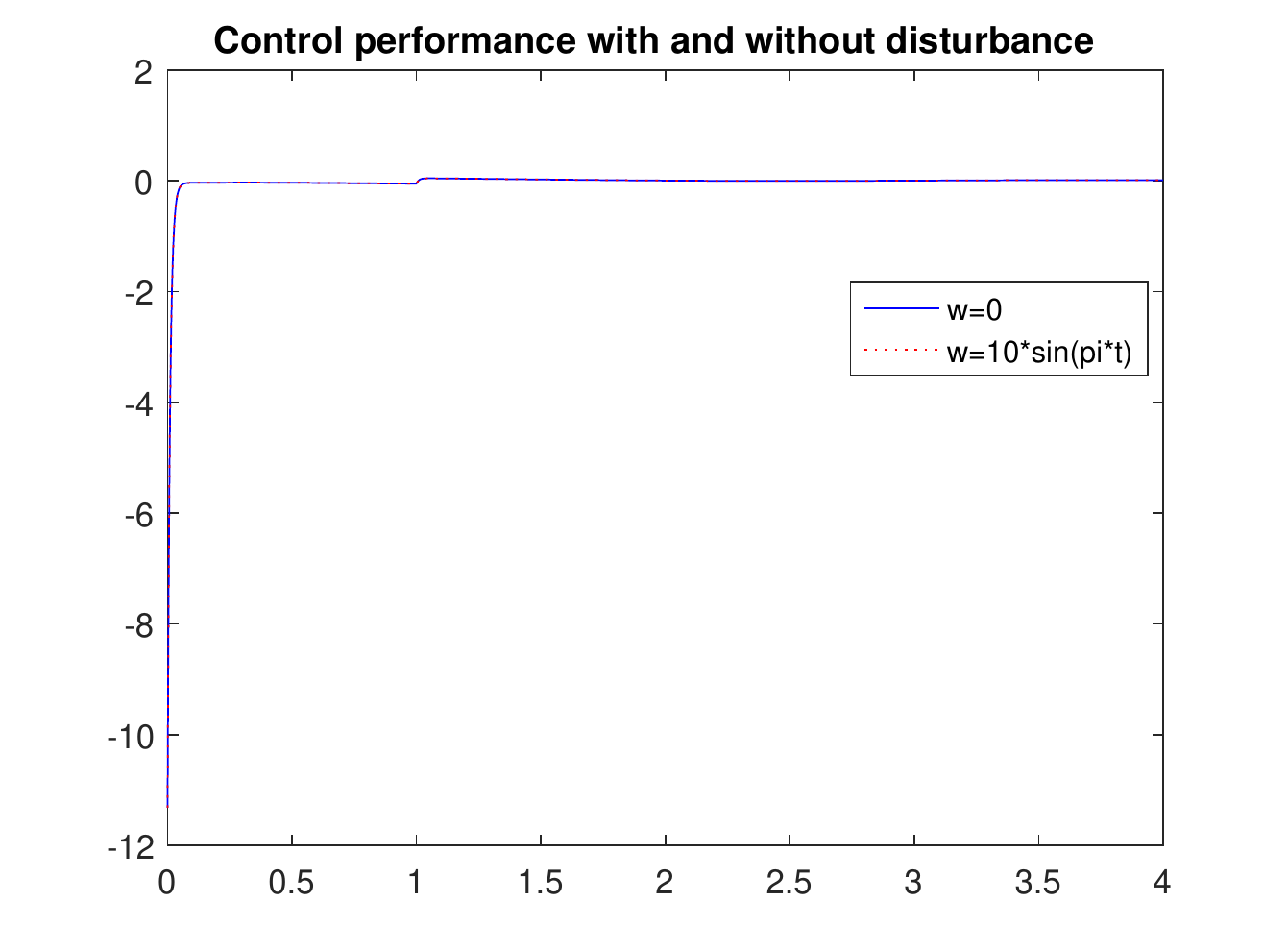}
\caption{The PID PDC behaviour with and without the disturbance }
\label{control_TORA}
\end{figure}

We set $\tau = 1$ for (\ref{if2}) in this example.
The minimal $\gamma=9.9$ is obtained by the bisection procedure. For this value of $\gamma$, Algorithm \ref{alg2}
needs $4$ iterations to arrive the following numerical values for implementing  PID PDC (\ref{if2}):
\[
\begin{array}{l}
R_{P1} = [-7.1101, -16.1981, 11.42817], \\
R_{P2} = [-5.5390, -11.9724, 8.5207],\\
R_{P3} = [-5.7119, -12.9499, 9.0553], \\
R_{P4} = [-5.7189, -12.9240, 9.0397],\\

R_{I1} = [-0.3471, -1.0139, 0.6820], \\
R_{I2} = [-0.3450, -1.01858, 0.6830],\\
R_{I3} = [-0.4091, -1.1184, 0.7669], \\
R_{I4} = [-0.4337, -1.1552, 0.7969],\\

R_{D1} = [0.8038, 2.0883, -1.2315], \\
R_{D2} = [0.6084, 1.3740, -0.7847],\\
R_{D3} = [0.5537, 1.4268, -0.7742], \\
R_{D4} = [0.5048, 1.4486, -0.7513].\\
\end{array}
\]
Figs. \ref{state_TORA}-\ref{control_TORA} respectively show the behavior of
system state and control with disturbance $w=10sin(\pi t)$ and with no disturbance.
The initial state condition is $x(0) = (0, 0, 0.5, 0)^T$.
The TORA system is smoothly stabilized well by PID PDC.

Algorithm \ref{alg1} achieves worse $\gamma=10.3$ and
needs $11$ iterations for converge for this value of $\gamma$. Fig.\ref{caseA_Alg1_Alg2_iteration} shows
the convergence behaviour of Algorithm \ref{alg1} (for $\gamma=10.3$ and Algorithm \ref{alg2}
(for $\gamma=9.9$).

\begin{figure}[h]
\centering
\includegraphics[width=0.8 \columnwidth]{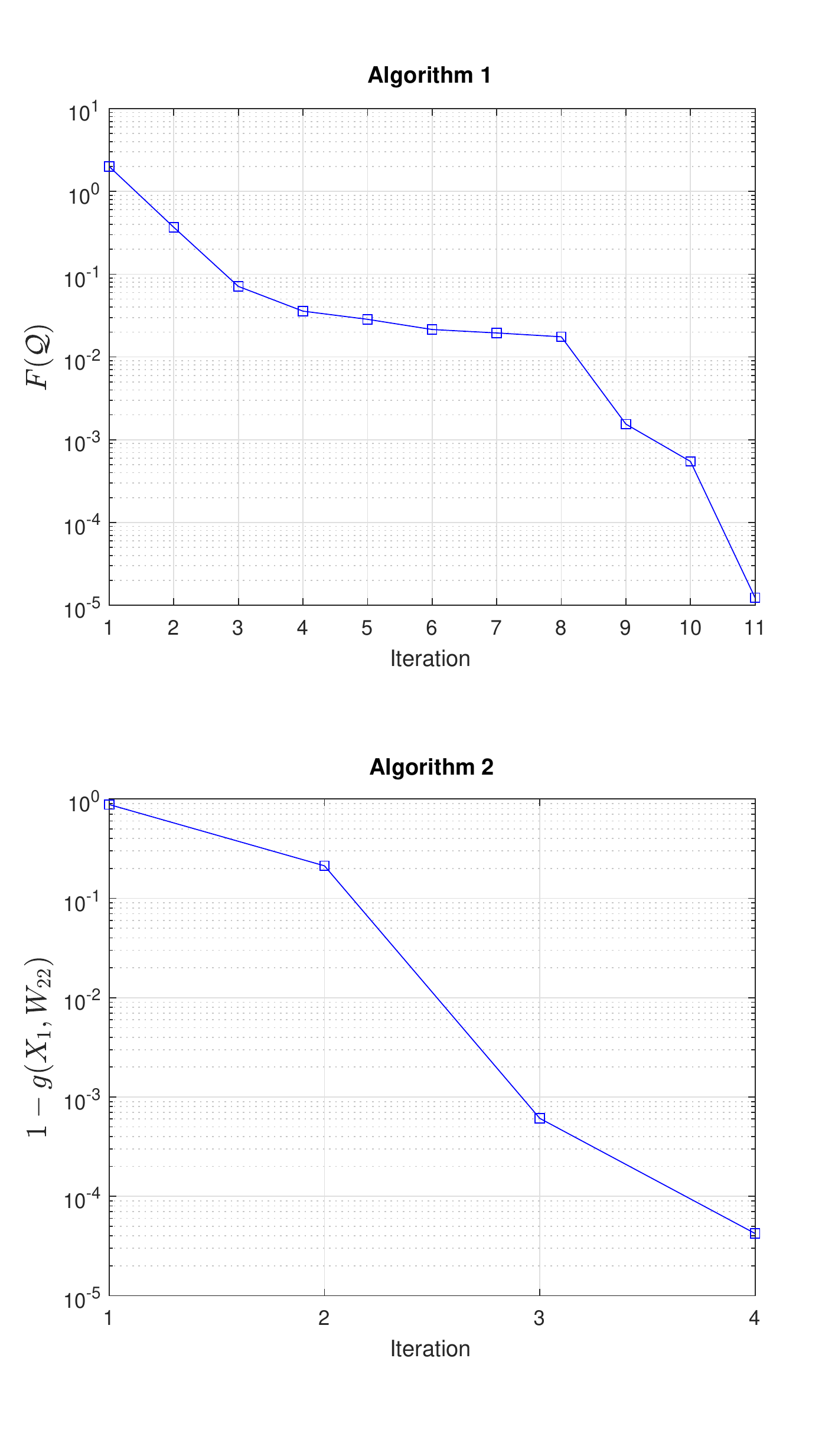}
\caption{Convergence performance by Algorithm \ref{alg1} and Algorithm \ref{alg2} for the TORA system}
\label{caseC_Alg1_Alg2_iteration}
\end{figure}

\section{Conclusion}
This paper has addressed the problem of designing $H_{\infty}$  PID PDC for T-S systems based on a
parameterized bilinear matrix inequality (PLMI), which is a system of infinitely many bilinear matrix
inequalities. Efficient computational procedures for this PLMI have been developed. Their merit has been
analysed through the benchmark examples. In the end, the effectiveness of PID PDC in smoothly stabilizing
nonlinear systems has been confirmed.

\bibliographystyle{ieeetr}
\bibliography{control}

\begin{thebibliography}{10}

\bibitem{TS85}
T.~Tagaki and M.~Sugeno, ``Fuzzy identification of systems and its applications
  to modeling and control,'' {\em IEEE Trans. Systems, Man, and Cybernetics},
  vol.~SMC-15, pp.~116--132, Jan 1985.

\bibitem{TANY01}
H.~D. Tuan, P.~Apkarian, T.~Narikiyo, and Y.~Yamamoto, ``Parameterized linear
  matrix inequality techniques in fuzzy control system design,'' {\em IEEE
  Trans. Fuzzy Syst.}, vol.~9, no.~2, pp.~324--332, 2001.

\bibitem{TANK04}
H.~D. Tuan, P.~Apkarian, T.~Narikiyo, and M.~Kanota, ``New fuzzy control model
  and dynamic output feedback paralell distributed compensation,'' {\em IEEE
  Trans. Fuzzy Syst.}, vol.~12, no.~2, pp.~13--21, 2004.

\bibitem{AT00}
P.~Apkarian and H.~D. Tuan, ``Parameterized linear matrix inequalities in
  control theory,'' {\em SIAM J. Control and Optimization}, vol.~38, no.~4,
  pp.~1241--1264, 2000.

\bibitem{ACL05}
K.~H. Ang, G.~Chong, and Y.~Li, ``{PID} control system analysis, design, and
  technology,'' {\em IEEE Trans. Control Syst. Tech.}, vol.~13, no.~4,
  pp.~559--576, 2005.

\bibitem{AT03}
M.~Araki and H.~Taguchi, ``Two-degree-of-freedom {PID} controller,'' {\em Int.
  J. of Control, Automation and Systems}, vol.~4, pp.~401--411, 2003.

\bibitem{GA11}
P.~Gahinet and P.~Apkarian, ``Structured {$H_{\infty}$} synthesis in
  {MATLAB},'' in {\em Proc. of IFAC 2011, Milan}, pp.~1--5, Jun. 2011.

\bibitem{GH15}
O.~Garpinger and T.~Hagglund, ``Software-based optimal {PID} design with
  robustness and noise sensitivity constraints,'' {\em J. of Process Control},
  vol.~33, pp.~90--101, 2015.

\bibitem{BHA15}
S.~Boyd, M.~Hast, and K.~J. Astrom, ``{MIMO PID} tuning via iterated {LMI}
  restriction,'' {\em Int. J. Robust and Nonlinear Control}, vol.~26,
  pp.~1718--1731, 2016.

\bibitem{HTN17}
S.~Hosoe, H.~D. Tuan, and T.~N. Nguyen, ``{2D} bilinear programming for robust
  {PID/DD} controller design,'' {\em Int. J. Robust Nonlinear Control},
  vol.~27, pp.~461--482, 2017.

\bibitem{GLCP15}
P.~Gil, C.~Lucena, A.~Cardoso, and L.~B. Palma, ``Gain tuning of fuzzy {PID}
  controllers for {MIMO} systems: A performance-driven approach,'' {\em IEEE
  Trans. Fuzzy Syst.}, vol.~23, no.~4, pp.~757--768, 2015.

\bibitem{ZWLH01}
F.~Zheng, Q.-G. Wang, T.~H. Lee, and X.~Huang, ``Robust {PI} controller design
  for nonlinear systems via fuzzy modeling approach,'' {\em IEEE Trans.
  Systems, Man, and Cybernetics-Part A: Systems and Humans}, vol.~31, no.~6,
  pp.~666--675, 2001.

\bibitem{CGLV16}
K.~Cao, X.~Gao, H.~K. Lam, and A.~Vasilakos, ``${H}_{\infty}$ fuzzy {PID}
  control synthesis for {Tagaki-Sugeno} fuzzy systems,'' {\em IET Control
  Theory \& Applications}, vol.~10, no.~6, pp.~607--616, 2016.

\bibitem{ANP08}
P.~Apkarian, D.~Noll, and O.~Prot, ``A trust region spectral bundle method for
  nonconvex eigenvalue optimization,'' {\em SIAM Journal on Optimization},
  vol.~19, no.~1, pp.~281--306, 2008.

\bibitem{Detal12}
Q.~T. Dinh, S.~Gumussoy, W.~Michiels, and M.~Diehl, ``Combining convex-concave
  decompositions and lienarization approaches for solving {BMIs}, with
  application to static outputfeedback,'' {\em IEEE Trans. Automat. Control},
  vol.~57, pp.~1377--1390, 2012.

\bibitem{BT97}
V.~Blondel and J.~N. Tsitsiklis, ``{NP}-hardness of some linear control design
  problems,'' {\em SIAM J. Control Optimiz.}, vol.~35, pp.~2118--2127, 1997.

\bibitem{AN06}
P.~Apkarian and D.~Noll, ``Nonsmooth ${H}_{\infty}$ synthesis,'' {\em IEEE
  Trans. Automatic Control}, vol.~51, pp.~71--86, Jan 2006.

\bibitem{ABN07}
P.~Apkarian, V.~Bompart, and D.~Noll, ``Non-smooth structured control design
  with application to {PID} loop-shaping of a process,'' {\em Int. J. Nonlinear
  Robust Control}, vol.~17, pp.~1320--1342, 2007.

\bibitem{RTN14}
U.~Rashid, H.~D. Tuan, and H.~H. Nguyen, ``Joint optimization of source
  precoding and relay beamforming in wireless {MIMO} relay networks,'' {\em
  IEEE Trans. on Commun.}, vol.~62, pp.~488--499, 2014.

\bibitem{PTKN12}
H.~A. Phan, H.~D. Tuan, H.~H. Kha, and D.~T. Ngo, ``Nonsmooth optimization for
  efficient beamforming in cognitive radio multicast transmission,'' {\em IEEE
  Trans. Signal Processing}, vol.~60, pp.~2941--2951, Jun. 2012.

\bibitem{Tuybook}
H.~Tuy, {\em Convex analysis and global optimization (second edition)}.
\newblock Springer, Berlin, 2016.

\bibitem{CRF96}
S.~G. Cao, N.~W. Rees, and G.~Feng, ``Stability analysis and design for a class
  of continuous-time fuzzy control systems,'' {\em Int'l. J. of Control},
  vol.~64, no.~6, pp.~1069--1087, 1996.

\bibitem{TW04}
K.~Tanaka and H.~O. Wang, {\em Fuzzy control systems design and analysis: a
  linear matrix inequality approach}.
\newblock John Wiley \& Sons, 2004.

\bibitem{TIW98}
K.~Tanaka, T.~Ikeda, and H.~O. Wang, ``Fuzzy regulators and fuzzy observers:
  relaxed stability conditions and {LMI}-based designs,'' {\em IEEE Trans.
  Fuzzy Syst.}, vol.~6, no.~2, pp.~250--265, 1998.

\end{thebibliography}
\end{document}